%% file: main.tex
\documentclass[10pt, conference, letterpaper]{IEEEtran}
\IEEEoverridecommandlockouts
\usepackage{graphicx}
\usepackage{balance}
\usepackage{comment}
\usepackage{graphics}
\usepackage{verbatim}
\usepackage{epsfig}
\usepackage{latexsym}
\usepackage{subfigure}
\usepackage{algorithm}
\usepackage{color}
\usepackage{amsmath}
\usepackage{tabularx}
\usepackage{multirow}
\usepackage[bottom]{footmisc}
\usepackage{algpseudocode}
\usepackage{amssymb}
\usepackage[english]{babel}
\usepackage{blindtext}
\usepackage{epstopdf}
\usepackage{cite}
\usepackage{url}

\begin{document}
\title{WiEat: Fine-grained Device-free Eating Monitoring Leveraging Wi-Fi Signals}

\author{Chen Wang, Zhenzhe Lin, Yucheng Xie, Xiaonan Guo, Yanzhi Ren, Yingying Chen 
	\thanks{Chen Wang is with the Rutgers University and Louisiana State University. This work was done during Chen Wang's Ph.D. study at Rutgers University. Email:chenwang1@lsu.edu}.
	\thanks{Zhenzhe Lin and Yingying Chen are with WINLAB, Rutgers University. Email:zhenzhe.lin@rutgers.edu, $yingche$@scarletmail.rutgers.edu. }
	\thanks{Yucheng Xie and Xiaonan Guo are with Indiana University-Purdue University Indianapolis. Email: $\left\{ {yx11, xg6} \right\}$@iupui.edu. }
	\thanks{Yanzhi Ren is with University of Electronic Science and Technology of China. Email: renyanzhi05@uestc.edu.cn. }
}

\maketitle

\input{./text_infocom/abstract.tex}

\input{./text_infocom/intro.tex}
\input{./text_infocom/related.tex}

\input{./text_infocom/system.tex}

\input{./text_infocom/methodology.tex}

\input{./text_infocom/methodology2.tex}
\input{./text_infocom/performance.tex}

\input{./text_infocom/discussion.tex}

\input{./text_infocom/conclusion.tex}


\bibliographystyle{../bib/IEEEtran}
\bibliography{./bib/main}

\end{document}

%% file: text_infocom/abstract.tex
\begin{abstract}
	Eating is a fundamental activity in people's daily life. Studies have shown that many health-related problems such as obesity, diabetes and anemia are closely associated with people's unhealthy eating habits (e.g., skipping meals, eating irregularly and overeating). Traditional eating monitoring solutions relying on self-reports remain an onerous task, while the recent trend requiring users to wear expensive dedicated hardware is still invasive. 
To overcome these limitations, in this paper, we develop a device-free eating monitoring system using WiFi-enabled devices (e.g., smartphone or laptop). 
Our system aims to automatically monitor users' eating activities through identifying the fine-grained eating motions and detecting the chewing and swallowing.
	In particular, our system extracts the fine-grained Channel State Information (CSI) from WiFi signals to distinguish eating from non-eating activities and further recognizing users' detailed eating motions with different utensils (e.g., using a folk, knife, spoon or bare hands). Moreover, the system has the capability of identifying chewing and swallowing through detecting users' minute facial muscle movements based on the derived CSI spectrogram. Such fine-grained eating monitoring results are beneficial to the understanding of the user's eating behaviors and can be used to estimate food intake types and amounts. Extensive experiments with $20$ users over $1600$-minute eating show that the proposed system can recognize the user's eating motions with up to $95\%$ accuracy and estimate the chewing and swallowing amount with $10\%$ percentage error.
\end{abstract}

%% file: text_infocom/intro.tex
\section{Introduction}

Eating, an essential activity for the energy intake and nutrition supply of a human, has been known to be closely related to people's health.
A surfeit of food could lead to the excess of calorie intake, gaining body weight and various health-related problems such as cardiovascular diseases, diabetes, stomach cancers~\cite{WHO2019consequence}.
Whereas the imbalanced or insufficient food intake could not fulfill the daily body needs and further result in nutritional deficiency problems such as anemia, osteoporosis and scurvy, which impedes the cell recovery and growth, especially for the patients, teenagers and seniors. 
The recent U.S. reports show that $70.2\%$ of American people suffer from overweight or obese problems~\cite{overweight2019} and $90\%$ of U.S. population have a nutrient deficiency~\cite{report2019}.
Moreover, recent surveys show that families in many regions of the world have become smaller, and there are more and more one-person households, besides, eating patterns have become irregular, informal, and individualized in the form of more eating alone~\cite{kwon2018eating}. By 2006, nearly 60 percent of Americans regularly ate on their own, according to the American Time Use Survey. Today, that number is even higher~\cite{washingtonpost2015}. Furthermore, research into eating alone is sparse, but some studies have suggested that eating alone may affect how much we eat, what we eat, and our mood~\cite{theguardian2018}. It is thus essential to provide a way for those people who are often eating alone to automatically detect and monitor their eating behaviors to help them keep a good dietary habit and also provide a tool for medical research.
Traditional eating monitoring methods mainly rely on self-reports (e.g., food logs~\cite{loewy2019diet} or food journals~\cite{yazio2019diet}). The smartphone Apps such as YouEat~\cite{feel2019} and Cara Care~\cite{your2019} allow users to manually record the food diaries via text logs or pictures. However, these self-report methods require user's participation to memorize or record the details in eating behaviors, and they all suffer from the subjective biases and the memory recall imprecision~\cite{hill2001validity}. 
The food scanner or the calorie calculator advance the self-monitoring methods by tracking the type of food and calculating the calories consumption~\cite{medical2017top}. 
But these methods usually require expensive dedicated hardware and the users' active participation is still inevitable. 

Recent years, the emerging mobile sensing technologies
have enabled several automatic eating monitoring systems such as smart utensil-based method~\cite{huang2018smart} and wearable-based methods~\cite{thomaz2015practical,bedri2017earbit,amft2010wearable}.
In particular, Smart-U develops a special spoon to recognize the types of food by its reflected light spectra~\cite{huang2018smart}. 
Thomaz et al.~\cite{thomaz2015practical} utilize the accelerometer on a smartwatch to capture the eating moments by recognizing the user’s hand motions during eating.
Amft et al.~\cite{amft2010wearable} detect the air-conducted vibrations of food chewing by a condenser microphone embedded in an ear pad to understand the food textures.
However, the smart utensil-based method is limited to a single utensil, while the wearable-based approaches can only detect partial body motions where the devices are worn (e.g., hand or jaw). They have limited capability to provide comprehensive eating monitoring. Moreover, these approaches all require additional dedicated devices. 




\begin{figure}
	\centering
	\includegraphics[width=3in]{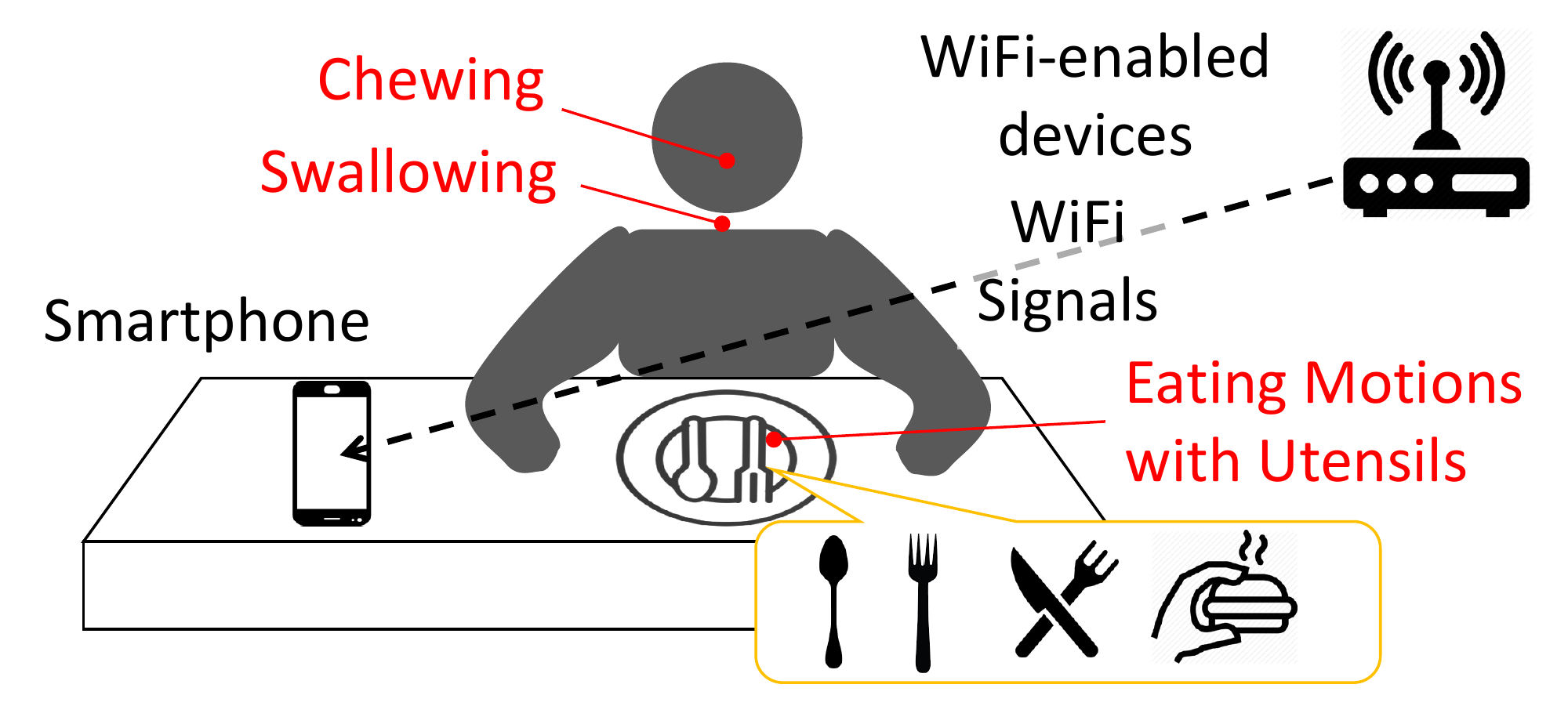}
	\vspace{-4 mm}
	\caption{An illustration of the proposed eating monitoring system.}
	\vspace{-5mm}
	\label{fig:intro_illustration}
\end{figure}

Different from the existing work, we aim at developing a personalized eating monitoring system to automatically provide comprehensive eating monitoring without active user participation. Figure~\ref{fig:intro_illustration} illustrates the basic idea of our approach. A user's smartphone (or an IoT device) is placed on a dinning table, which utilizes the WiFi signals received from a WiFi-enabled devices (e.g., Access point, laptop) to provide automatic eating monitoring.
The proposed system could not only track users' eating motions but also detect the users' mastication muscle movements, which are critical to infer users' eating behaviors and estimate the food type and amount. 
Although using WiFi signal to recognize human activities has shown its initial success, such as location-oriented activity identification~\cite{wang2014eyes}, the fitness assistance~\cite{guo2018device} and vital sign monitoring (e.g., breathing rate)~\cite{liu2015tracking},
it cannot be directly used for eating monitoring, which requires capturing both the relatively large scale hand motions and the minute facial muscle movements during chewing and swallowing.
To realize a WiFi-enabled eating monitoring system, several challenges need to be addressed:
1) It is not trivial to recognize the eating activities from the many other different human activities, since people conduct many different indoor activities besides eating in a day;
2) The various eating motions with different utensils (e.g., fork, spoon, knife and bare hand) all involve similar hand movements (i.e., delivering foods from a plate to mouth) and thus further distinguishing these eating motions based on the noisy WiFi signals is challenging;
3) The chewing and swallowing only exhibit minute facial muscle movements, which are hard to be captured from the WiFi signals; 
4) Because the smartphone is embedded with small internal WiFi antennas, the quality of the received WiFi signals is much lower than the devices using external antennas, and how the smartphone could provide WiFi sensing is still unexplored. 

To address these challenges, we develop WiEat, a system that leverages the channel state information (CSI) extracted from smartphones or WiFi-enabled IoT devices to monitor a user's fine-grained daily eating activities. In particular, the proposed system develops a cluster-based method to differentiate the eating motions from the many other non-eating activities by capturing the unique physiological characteristics of eating motions. We then propose to extract the unique spectrogram features of eating motions and develop a soft decision-based algorithm to further recognize how a user eats (i.e., with which utensil).
Moreover, we utilize a Minute Motion Reconstruction method to capture the minute facial muscle movements of chewing and swallowing and develop an accumulated power spectral density method to detect the periods of these minute motions for deriving the statistics of chewing and swallowing.

\noindent
Our contributions are summarized as follows:
\begin{itemize}
	\item We demonstrate that the CSI extracted from a smartphone or an IoT device can be used to provide fine-grained eating monitoring, which not only recognizes the eating motions with different utensils but also capture the minute muscle movements of chewing and swallowing. 
	
	\item We develop a device-free eating monitoring system based on CSI to automatically track people's eating activity, which can be easily deployed on the user's smartphone or an IoT device without incurring additional costs.
	
	\item We derive the CSI spectrogram and develop a soft decision-based approach to identify the various eating motions according to the utensils the user uses. Moreover, we capture the minute facial muscle movements based on minute chewing/swallowing motion reconstruction and develop the accumulated power spectral density method to derive the chewing and swallowing statistics.

	\item Extensive experiments with $20$ people over $1600$-minute eating show that our system can recognize the user's eating motions and estimate the fine-grained chewing and swallowing statistics with high accuracy.
\end{itemize}

%% file: text_infocom/related.tex
\section{Related Work}

Traditional eating monitoring methods are mainly based on questionnaires or self reports~\cite{loewy2019diet, yazio2019diet, fallaize2014online}.
Fallaize~\textit{et al.}~\cite{fallaize2014online} design Food4Me, an online Food Frequency Questionnaires (FFQ) system, to collect a user's nutrient intake data.
The recent smartphones Apps~\cite{loewy2019diet, yazio2019diet} enable the user to conduct self reports with more flexibility and convenience. For example,
My Macros+~\cite{loewy2019diet} provides a diet tracking solution by asking the user to manually log and select what he/she eats.
However, these methods require the user's active participation and suffer from the subjective bias. They are also subject to the user's memory recall imprecisions. The food scanner or the calorie calculator increases the accuracy to recognize foods and calculate the calories~\cite{medical2017top,tellspc2019building,scio2019scan}, but they still rely on users' participation and the expensive dedicated hardware. 




To reduce the user's efforts, some vision-based methods are developed to automatically monitor the user's eating using cameras ~\cite{kong2012dietcam, zhu2010image}. 
DietCam~\cite{kong2012dietcam} performs automatic dietary assessment by processing the photo strings or short videos taken by the user's mobile device, while eButton~\cite{sun2014ebutton} relies on a camera attached to the chest location to capture and evaluate the diet. However, the vision-based approaches may raise privacy concerns, because images often capture the user's sensitive information (e.g., eating with whom and where). Instead, there are some studies focusing on developing smart utensils to analyze the food intake automatically. CogKnife~\cite{kojima2016cogknife} attaches a small microphone to a knife to identify food types based on the recorded food cutting sounds. Smart-U~\cite{huang2018smart} uses a dedicated spoon equipped with a LED light \& sensor of recognizing different foods based on their reflected light spectra. But these methods limited to a dedicated spoon or knife are hard to provide the comprehensive eating monitoring, where people could eat flexibly with other utensils or bare hands. 




There are also active studies using the wearable devices (e.g., head-worn and wrist-worn devices) to provide automatic eating monitoring~\cite{thomaz2015practical, bedri2017earbit,amft2010wearable, amft2005analysis}.
Thomaz~\textit{et al.}~\cite{thomaz2015practical} utilize the accelerometer on a smartwatch to capture the eating moments by recognizing the user's hand motions during eating. Bedri~\textit{et al.}~\cite{bedri2017earbit} develop an eating episode detection system based on a dedicated ear-worn device, which is embedded with an inertial sensor behind the ear to detect people's chewing motions. Similarly, Amft~\textit{et al.}~\cite{amft2010wearable} detect the air-conducted vibrations of food chewing by a condenser microphone embedded in an ear pad to understand the food textures. Because a single wearable device limited by its body-location is hard to provide comprehensive eating monitoring, the studies~\cite{sazonov2008non,ye2015automatic} propose to exploit multiple wearable devices. Sazonov~\textit{et al.}~\cite{sazonov2008non} develop a system based on multiple head-worn devices (e.g., a piezoelectric sensor below the outer ear and a throat microphone) to detect the chewing and swallowing. Ye~\textit{et al.}~\cite{ye2015automatic} apply a head-worn and a wrist-worn devices to detect both the hand-to-mouth eating gestures and the chewing behaviors based on their the accelerometers.
However, all the above methods are intrusive to the user by requiring the user to wear one or multiple dedicated devices. 

Recent years, the WiFi sensing has shown the initial success to provide the non-invasive human activity recognition~\cite{wang2014eyes, guo2018device, liu2015tracking}.
E-eyes~\cite{wang2014eyes} utilizes the WiFi signals to provide device-free location-oriented human activity identification. 
Guo~\textit{et al.}~\cite{guo2018device} use the WiFi signals from IoT devices to provide device-free fitness assistance.
Liu~\textit{et al.}~\cite{liu2015tracking} develops a vital sign tracking system, which can detect minute human body motions like breathing via WiFi signals.
Thus, in this paper, we propose to utilize the WiFi signals from the user's smartphone or IoT device to provide fine-grained eating monitoring. 



%% file: text_infocom/system.tex
\section{System}



\subsection{\textbf{What Need to be Monitored for Analyzing Eating?}}
Understanding the daily dietary and eating behaviors is critical for an individual to maintain healthy eating. Based on that the users could know whether to reduce eating for bodyweight management or increase the intake of some types of food for obtaining sufficient minerals or vitamins. The purpose of a comprehensive eating monitoring is to understand when the users eat, what they eat and how much they eat? To achieve this, we need to capture the eating activities, recognize the detailed eating motions (e.g., with which utensil) and monitor the chewing and swallowing. 

\noindent
\textbf{Capturing Eating Activities. }People conduct many different indoor activities around the table in a day, such as working in front of a laptop, doing exercises, talking with people, reading books, watching TV and eating. To provide fine-grained eating monitoring, the first step is to detect the user eating activities and differentiate them from the many other activities. Based on that we could capture the eating moments and infer the starting time, ending time and the duration, which is important for inferring when the user eat (e.g., the user's eating behavior pattern to have regular meals or skip meals). Moreover, by setting the boundaries of the eating moments, we could further derive the detailed eating motions within the eating periods without processing all the WiFi data. 


\noindent
\textbf{Recognizing Eating Motions (with Utensils). }
To further understand what they eat, we need to recognize the detailed eating motions (i.e., with which utensil the user delivers the food to mouth). This is because the utensils used by the user could reflect the type of food they eat. Usually, a user may use a fork, knife, spoon or bard hands to eat, which are some commonly used utensils by over 4.5 billion people~\cite{eating2019}. According to a recent survey~\cite{buzzfeed2019}, $95\%$ people use a spoon to eat oatmeal and porridge; $70\%$ to $80\%$ use a fork to eat salad, macaroni and mashed potatoes; $90\%$ people use their bare hands to eat pizza, burger and French fries. And people are more likely to use a knife and a fork to eat steak and fish. Therefore, by recognizing the utensils used, we could estimate the type of food the user eats, which is important to evaluate the users' nutrition balance. Moreover, the differentiated detailed eating motions is helpful to count the hand-to-mouth delivering motions under each utensil category for estimating how much of each food the user eats.

\noindent
\textbf{Monitoring Chewing and Swallowing.} But the user's eating motions only provide coarse-grained information, a comprehensive eating monitoring system also need to track the user's chewing and swallowing, which show more detailed information about what the users eat and how much they eat. In particular, the number of chewing required per mouthful is related to the food texture and density~\cite{healthline2019}. For example, the high-density foods(e.g., steak and nuts) require up to chews, while the foods like fruit and vegetables require fewer chews (e.g., 10 to 15) to break down. The fluid foods (e.g., porridge) require several chews or no chews. Therefore, the number of chews together with the utensil the user uses provide more information for estimating the type of the food intake. Moreover, researches have pointed out that chews and swallows are two critical measurements for quantifying the food intake amount~\cite{sazonov2009toward,farooq2016automatic}. Sazonov~\textit{et al.} develop a model to estimate the amount of food intake ( in ounce) based on the chew counts and swallow counts. To take one step further, the food type information can be integrated into this model to further estimate the energy intake (i.e., in calorie)~\cite{farooq2016automatic}. Therefore, our system needs to capture the user's chewing and swallowing. 

\begin{figure}
	\centering
	\includegraphics[width=3.6in]{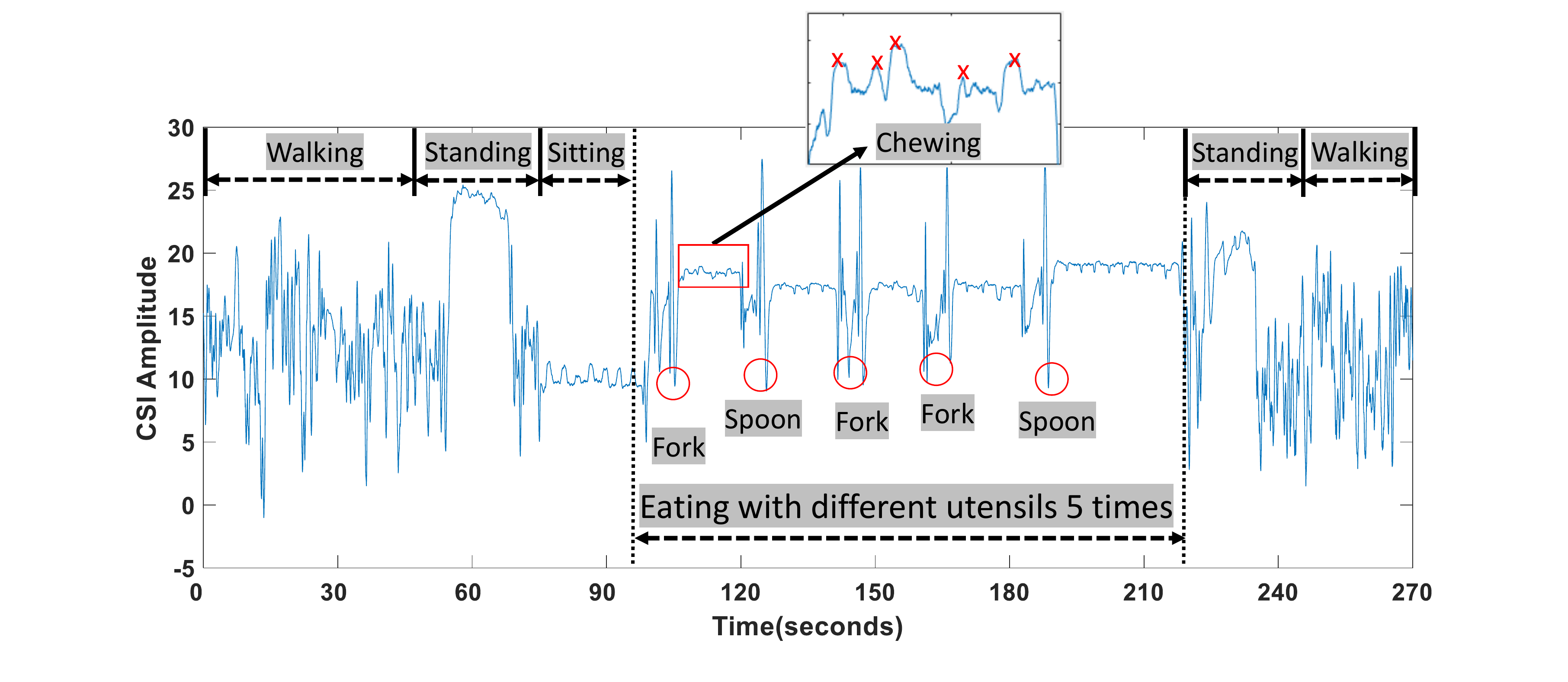}
	\vspace{-6mm}
	\caption{CSI amplitudes of one subcarrier under different human activities.}
	\vspace{-5mm}
	\label{fig:PreliminaryCSI}
\end{figure}

\subsection{\textbf{Feasibility Study}}

To provide the fine-grained eating monitoring, our basic idea is to extract the Channel State Information (CSI) from the user's smartphone or IoT device to capture both the user's detailed eating motions and the minute facial muscle movements of chewing and swallowing. CSI is a fine-grained measurement of the wireless channel in 30 subcarriers, with each subcarrier measures the state of a subchannel with the amplitude and phase information. Compared to single value of the traditional received signal strength, the CSI containing 30 subcarriers provide much more information to describe the propagation of the wireless signal, including its fading, scattering, multipath and the wireless interference. Thus, when a user is present in the signal propagation paths, his/her body motions could cause inferences to the WiFi signals in the form of reflection, absorption and refraction, which can be captured by the CSI. But a user's eating activity is very complicated, which include the large scale hand motions to deliver to the food to mouth and the minute muscle level jaw movements and pharynx movements to break down and ingest the food. Besides, a user can flexibly use different utensils to eat, which cause the eating recognition to be harder. 

In order to understand how the user's eating activity could incur the changes in the CSI, we conduct a preliminary study. In particular, we ask a participant to perform both eating and non-eating indoor activities, including walking, standing, sitting, eating with a fork and eating with a spoon, when the participant's smartphone is placed on the table as illustrated in Figure~\ref{fig:intro_illustration}. The CSI is collected from the smartphone for analysis. Figure~\ref{fig:PreliminaryCSI} shows the CSI amplitude of one subcarrier, where we mark the ground-truth of the participant's activities. Given that eating is a process when people deliver the foods to mouth multiple times, we observe such repetitive eating motions on the CSI amplitudes, which contain the user's eating with a fork or a spoon for five times. The eating motions also show different curve patterns compared to the non-eating activities. Moreover, after each eating motion (i.e., food delivery), we could find some slight vibrations (e.g., marked by the red rectangular), which are caused by the minute jaw movements of chewing. However, we also find that it's hard to further differentiate the eating motion using a fork from that using a spoon from the CSI amplitude. Furthermore, the chewing and swallowing motions are in small scale and easy to be submerged by noises. 

\subsection{System Design}
\begin{figure}[t]
	\centering
	\includegraphics[width=0.95\columnwidth]{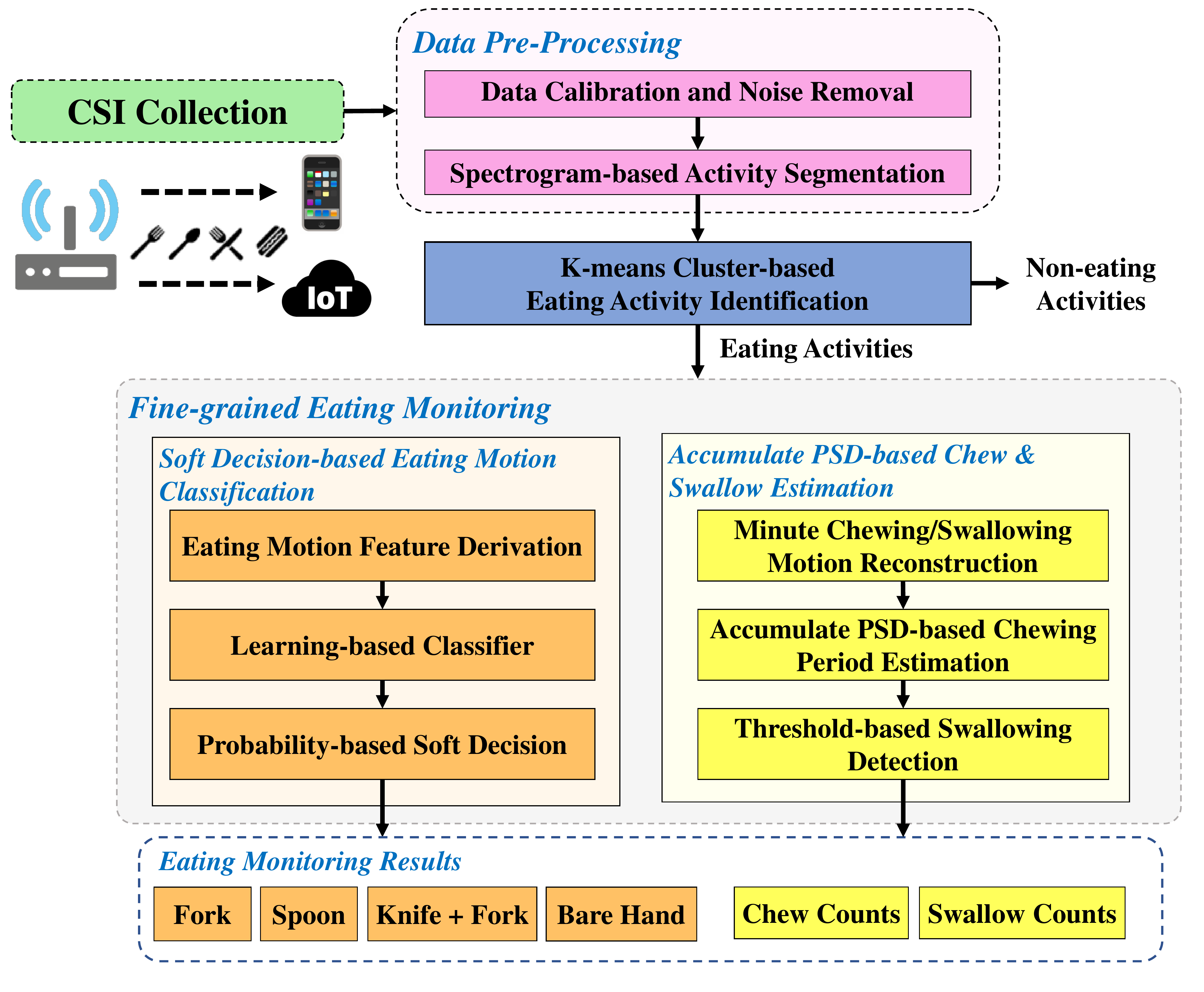}
	\vspace{-4mm}
	\caption{System Flow of WiEat}
	\vspace{-5mm}
	\label{fig:SystemFlow}
\end{figure}

The system flow of WiEat is shown in Figure~\ref{fig:SystemFlow}, which takes the CSI measurements extracted from the user's WiFi devices (e.g., smartphone and IoT device) as the input. \textit{Data Pre-processing} is first performed to calibrate the CSI data and segment the data based on the user's activities. In particular, \textit{Data Calibration and Noise Removal} removes the outliers and applies a bandpass filter to reduce the high frequency noises. \textit{Spectrogram-based Activity Segmentation} derives the spectrograms from the CSI measurements, based on which the short-time-energy is accumulated to determine the starting and ending time of the user's activities. 
With the user's various activities, \textit{K-means Cluster-based Eating Activity Identification} captures the unique characteristics of the eating activities from the CSI and utilizes K-means to differentiate the eating activities from the many various non-eating activities by separating them into different clusters. 
The core of our system for providing the fine-grained eating monitoring consists of two components, \textit{Soft Decision-based Eating Motion Identification} and \textit{Chewing and Swallowing Estimation}. Given the differentiated eating activities, \textit{Soft Decision-based Eating Motion Identification} further distinguish them to know the details of the user's eating motions (i.e., with which utensil the user eats). It performs \textit{Eating Motion Feature Extraction} to derive the statistic features from each CSI of subcarrier to capture the unique inherent behavioral characteristics of the user's different eating motions. The learning-based classifier recognizes an eating motion based on each CSI subcarrier data separately and calculate the decision probability of classifying the eating motion into every pre-trained categories (i.e., fork, spoon, knife + fork and bare hand). 
\textit{Probability-based Soft Decision} integrates the decision results from different CSI subcarriers by adding the category probabilities among all the CSI subcarriers, and the resulted greatest category probability corresponds to the final decision result. This method leverages the information from all the CSI subcarriers efficiently rather than only selecting some better CSI subcarriers and losing the information from the others.  

As the second major component, the \textit{Eating Motion Identification}, aims to quantify the chewing and swallowing motions by deriving the chewing period and measuring the chew count and swallow count. In particular, after each identified eating motion (i.e., food delivery), \textit{Minute Motion Reconstruction} is performed to magnify the small facial muscle movements of chewing and swallowing by reconstructing the such minute motion information from all the CSI subcarriers. To estimate the chewing period, we develop the \textit{Accumulate PSD-based Chew Detection} to analyze the repetitive patterns of the chew motions from the CSI power spectral density accumulated over all the CSI subcarriers in the frequency domain. To detect the swallowing, the \textit{Threshold-based Swallowing Detection} recognizes the swallowing motions by capturing the inherent muscle movement differences between chewing and swallowing based on amplitude range and peak-to-valley time interval. Then the statistics of the chewing and swallowing, including the chew count and swallow count can be obtained.

%% file: text_infocom/methodology.tex
\section{Fine-grained Eating Monitoring}



\subsection{Data Pre-processing}

\subsubsection{Data Calibration and Noise Removal} 
The WiFi signals suffer from severe wireless interferences and environmental noises in the daily environments (e.g., home, dining room and restaurant). We first perform data calibration to remove the outliers from the WiFi data and apply a band-pass filter with a sliding window to mitigating the interference of high frequency and low frequency noises. Moreover, we observe that the noise source came from environment-related changes usually present on a fixed frequency range, we then analyze the environment and define a empirical threshold to identify these noises from user activities and remove them.


\subsubsection{Spectrogram-based Activity Segmentation}
\begin{figure}[t]
	\centering
	\includegraphics[width=0.9\columnwidth]{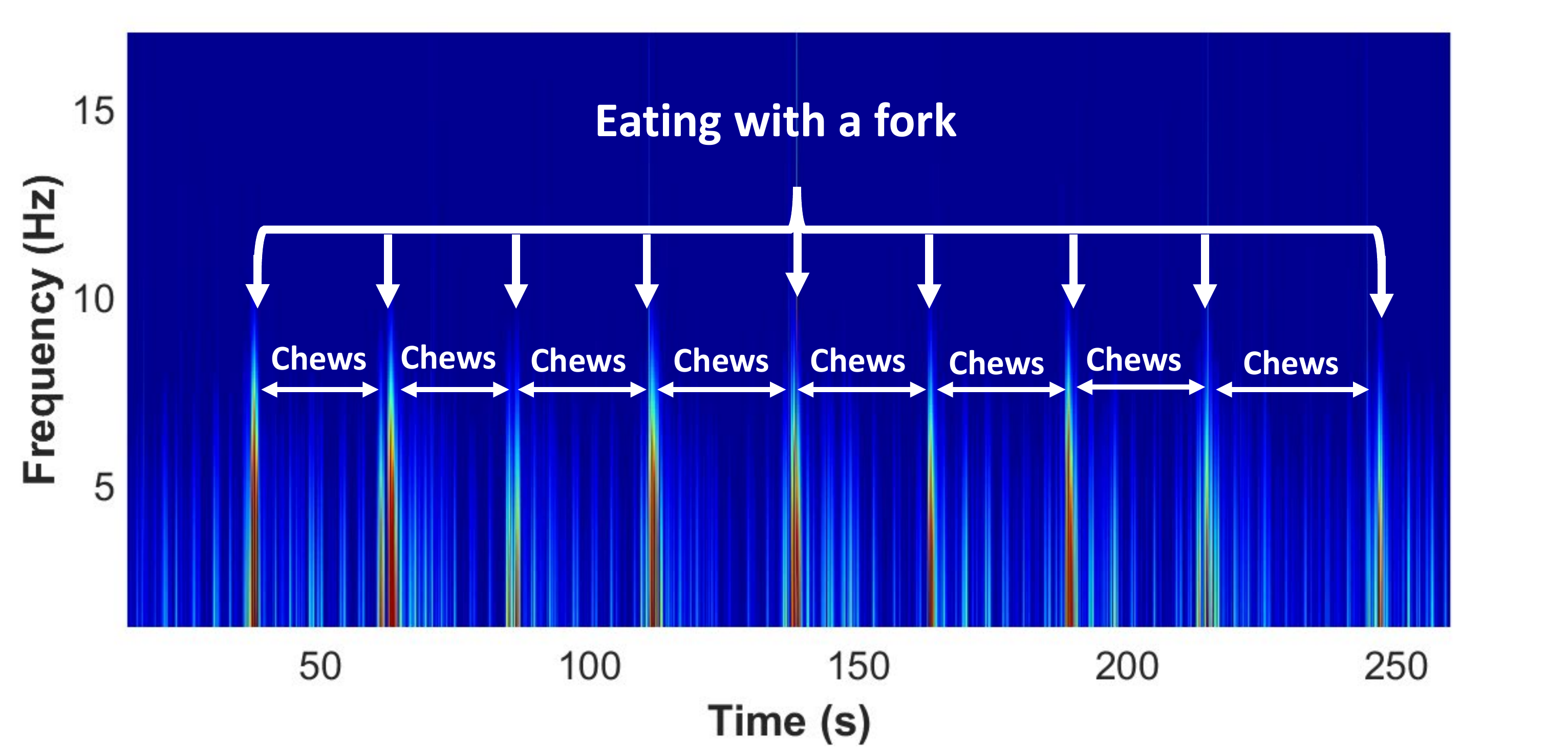}
	\vspace{-2mm}
	\caption{Illustration of spectrogram-based activity segmentation. }
	\vspace{-5mm}
	\label{fig:spectrogram}
\end{figure}

With the calibrated data, we utilize the spectrogram-based activity segmentation to detect the activities on the CSI data. We first apply with a moving variance filter to removes high-frequency noise that is unlikely to be caused by the corresponding human activities. Moreover, we using the cumulative power spectral density (CPSD) to calculate the integrated frequency-domain effected by human activities. Due to the large body movements always lasting for a short period and perform a ephemeral impulse in the frequency-domain, we need to reconstruct the CSI complex to enlarge the trifling variances into a distinct pattern. Inspired by the motivation, we adopt Cumulative Short Time Energy (CSTE) to capture each eating activity. Specifically, we calculate the CPSD by accumulating all the power spectral density along the frequency dimension in the corresponding spectrogram. The Cumulative Short Time Energy is then calculated based on the CPSD by the following formula to segment the eating activities with higher accuracy:

\vspace{-2mm}
\begin{equation}
\footnotesize
STE = \sum_{i=-\infty}^\infty[CPSD(i)W(n-i)]^2,
\vspace{-2mm}
\end{equation}

where $CPSD(i)$ is the cumulative power spectral density, $W(n)$ denotes the window function and $n$ represents the frame shift of samples.In the relative spectrogram in frequency domain shown in Figure~\ref{fig:spectrogram} we can observe clearer repetitive patterns, where the color degree of the wave represents for the strength of power amplitude. 

Based on this observation, we then use the zero energy points to segment the large activities. After recognizing all the peaks in the Figure~\ref{fig:spectrogram}, we find the two zero points at the left and right respectively of the peak. The corresponding points of the zero points marked in the figure, determine the staring and ending time of the activities caused by users, which could be used to segment the one completed activities.



\subsection{Differentiate Eating Activities from Non-eating}

In this section, we focus on differentiating the eating activities from non-eating indoor activities with our K-means cluster-based eating activities identification method. 

\begin{figure}[t]
	\centering
	\includegraphics[width=0.7\columnwidth]{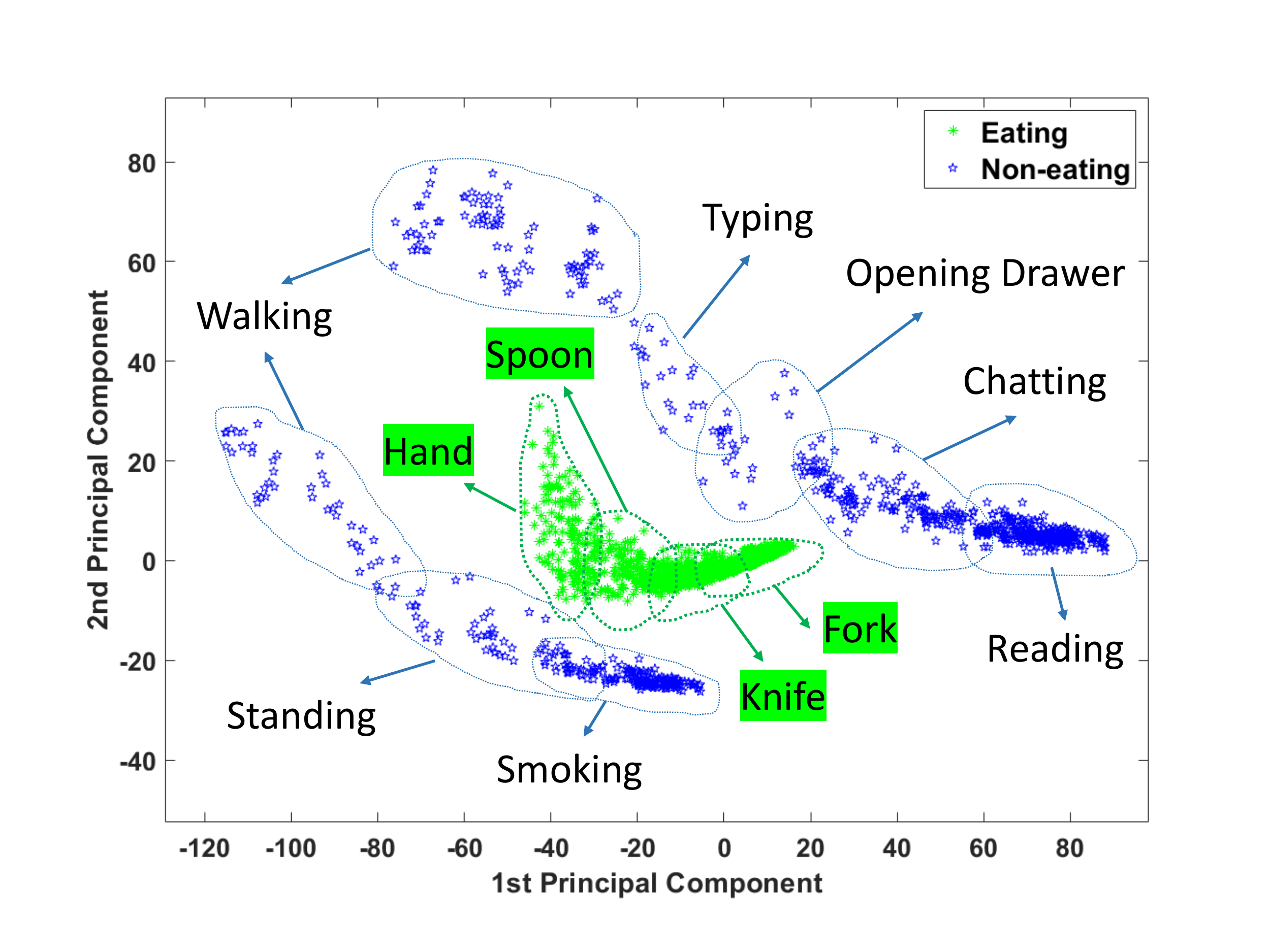}
	\vspace{-2mm}
	\caption{Clusters of eating activities and non-eating activities. }
	\vspace{-5mm}
	\label{fig:cluster}
\end{figure}

The eating activities are defined as the movement caused by people who deliver food from cutlery to mouth, showing the differences compare with the other non-eating activities caused by people who involved large body movements (e.g., walking, sitting and standing). The basic idea is to capture the wireless channel variances of all activities and then use the cluster-based method to differentiate them. We first examine how different activities influence the CSI complex. As shown in Figure~\ref{fig:cluster}, the eating activities and non-eating indoor activities could be highly differentiated into two clusters with two principal components through the principal component analysis (PCA). We observed that eating activities (i.e., using forks, using knifes, using spoons and using bare hands) are gathered in the middle position whether other non-eating activities such as reading, chatting and typing activities surround in the external of eating activities. This is because the eating activities presented as the repetitive patterns from hand to mouth, which have highly similarity to each other. Some people may argue that a few activities also include repetitive patterns from hand to mouth, i.e., smoking. Since more and more  countries would ban smoking in all indoor places, our system does not consider those activities as a distraction to our method.

Inspired by the above observations, we propose to use a cluster-based method to differentiate eating from non-eating activities. Specifically, K-means method is used to partition all the activities into two clusters with two geometric centroids calculated as $\mu _{1}$ and $\mu _{2}$. After select the distance between the eating activity centroid and the testing activity $\mu _{2}$ for cluster-based detection which denotes as $D_{c}=\left \|\mu _{1}-\mu _{2}  \right \|.$ Through the analysis above, we show that the cluster-based method has the capability of identifying the eating activities by applying the threshold $\xi$ to the $D_{c}$ as follows:

\vspace{-2mm}
\begin{equation}
\footnotesize
\begin{cases}
D_{c}\leqslant  \xi,  \ eating\ motions \\ 
D_{c}> \xi, \ non-eating\ motions 
\end{cases}
\vspace{-3mm}
\end{equation}

\subsection{Soft Decision-based Eating Motion Classification}

After detecting the eating activities, we further recognize the detailed eating motions according to which utensils the user uses, which provide us with the information about what the user eats and how much he/she eats. 

\subsubsection{Eating Motion Feature Derivation} 

To further classify the eating motion with utensils, we need to derive a series reliable features extracted from the CSI readings. Additionally, unique features of eating motions could eliminate the environmental noises in terms of WiFi signals suffer from the ambient interference. Based on our preliminary experimental investigations and detailed analysis of extracted CSI, we particularly choose $14$ features extracted from both time domain (e.g., \textit{root mean square}, \textit{average rectified value}, \textit{25th percentile value}, etc) and frequency domain (e.g., \textit{phase}, \textit{energy}, etc) of each sub-carrier.

\subsubsection{Learning-based Classifier} 
We adopt the learning-based classifier to further identify the eating motions with utensils. We use the Support Vector Machine (SVM) implemented by LIBSVM~\cite{chang2011libsvm} with linear kernel for building the classifier. Specifically, for each segmented CSI raw reading, we extract a set of eight features from all thirty subcarriers and then derive a two-dimensional with $30 \times 8$ vectors as the input for learning-based classifier. We denoted the two-dimensional vectors from selected thirty subcarriers as $v= [\upsilon_{1},...,\upsilon_{i},...,\upsilon_{30}]$, where $\upsilon_{i}$ includes the eight features mentioned above. Four prediction probabilities regarding different eating motions with utensils are defined as $\rho_{f},\rho_{k},\rho_{s},\rho_{h}$, corresponding to forks, knifes, spoons, hands, respectively. Then the estimated prediction probabilities for four different eating motions with utensils from the $i^{th}$ subcarrier can be obtained by using the following equation~\ref{eq: probality decision}:

\vspace{-3mm}
\begin{equation}
\footnotesize
\begin{cases}
P^{i} = \max \{ {\rho^{i}_{f},\rho^{i}_{k},\rho^{i}_{s},\rho^{i}_{h}} \}
\\
P_{total}= \sum_{i=1}^{30}[\rho^{i}_{f} + \rho^{i}_{k} + \rho^{i}_{s} + \rho^{i}_{h}]
\\

\end{cases}
\vspace{-1mm}
\label{eq: probality decision}
\end{equation}

\begin{figure}[t]
	\centering
	\includegraphics[width=0.55\columnwidth]{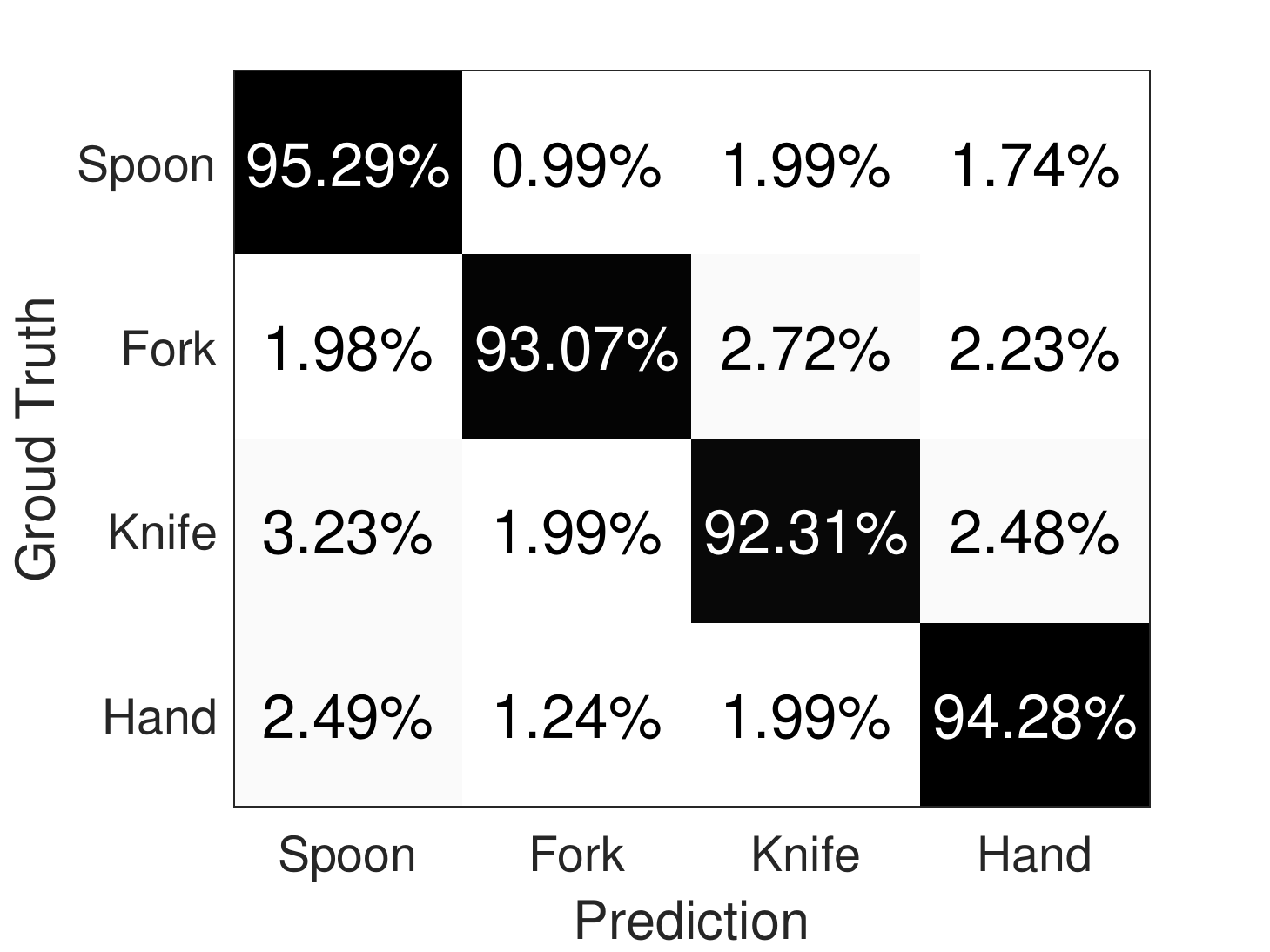}
	\vspace{-2mm}
	\caption{Confusion matrix of using soft decision-based classification to differentiating eating motions.}
	\vspace{-5mm}
	\label{fig:confusion_matrix_smartphone}
\end{figure}

\subsubsection{Probability-based Soft Decision Strategy}

Even though some carriers show lower sensitivity to the user's eating motions, they still contribute useful information to the classification decision. The traditional methods by majority vote over the hard decision results (i.e., one of the category) or subcarrier selection is hard to utilize the useful information from all the CSI subcarriers. 
Different from these methods, we develop a new probability-based soft decision strategy to leverage all the subcarriers and infer the eating motions with various utensils. In particular, the probability of classifying the eating motion to a utensil category based on each CSI subcarrier can be integrated with an assigned weight. The integrated probabilities of the all utensil category are compared and the utensil category with the largest integrated probability is the final decision. The soft decision-based eating motion classification can be described as:
\vspace{-2mm}
\begin{equation}
\footnotesize
argmax \left(\sum_{i=1}^{30} [p_f^{i} \cdot w_{f}^{i}], \sum_{i=1}^{30} [p_k^{i} \cdot w_{k}^{i}], \sum_{i=1}^{30} [p_s^{i} \cdot w_{s}^{i}], \sum_{i=1}^{30} [p_h^{i} \cdot w_{h}^{i}] \right),
\vspace{-2mm}
\end{equation}
where the assigned weight value $w$ is determined based on the variance of the CSI at each subcarrier, with the larger variance showing the higher sensitivity to the eating motions.
 
Figure~\ref {fig:confusion_matrix_smartphone} shows confusion matrix when using our probability-based soft decision strategy, where the columns of the confusion matrix denote the ground truth of an activity (e.g., eating with a spoon), while the rows represent the classified activity in our system. Each entry in the matrix represents the percentage of the correctly classified activity. The results show that our system can achieve over 90\% accuracy under the probability-based soft decision strategy. This is because the soft decision strategy regarding the prediction probability of eating motions in each subcarrier will be in the interval range between 0 and 1. Rather than hard-decision based method whose decision is only be 0 or 1, the probability-based soft decision strategy shows the insight which could provide the users with a more accurate classification results.

%% file: text_infocom/methodology2.tex
\begin{figure}[t]
	\centering
		\begin{minipage}{0.45\columnwidth}
		\centering
		\includegraphics[width=1.6in]{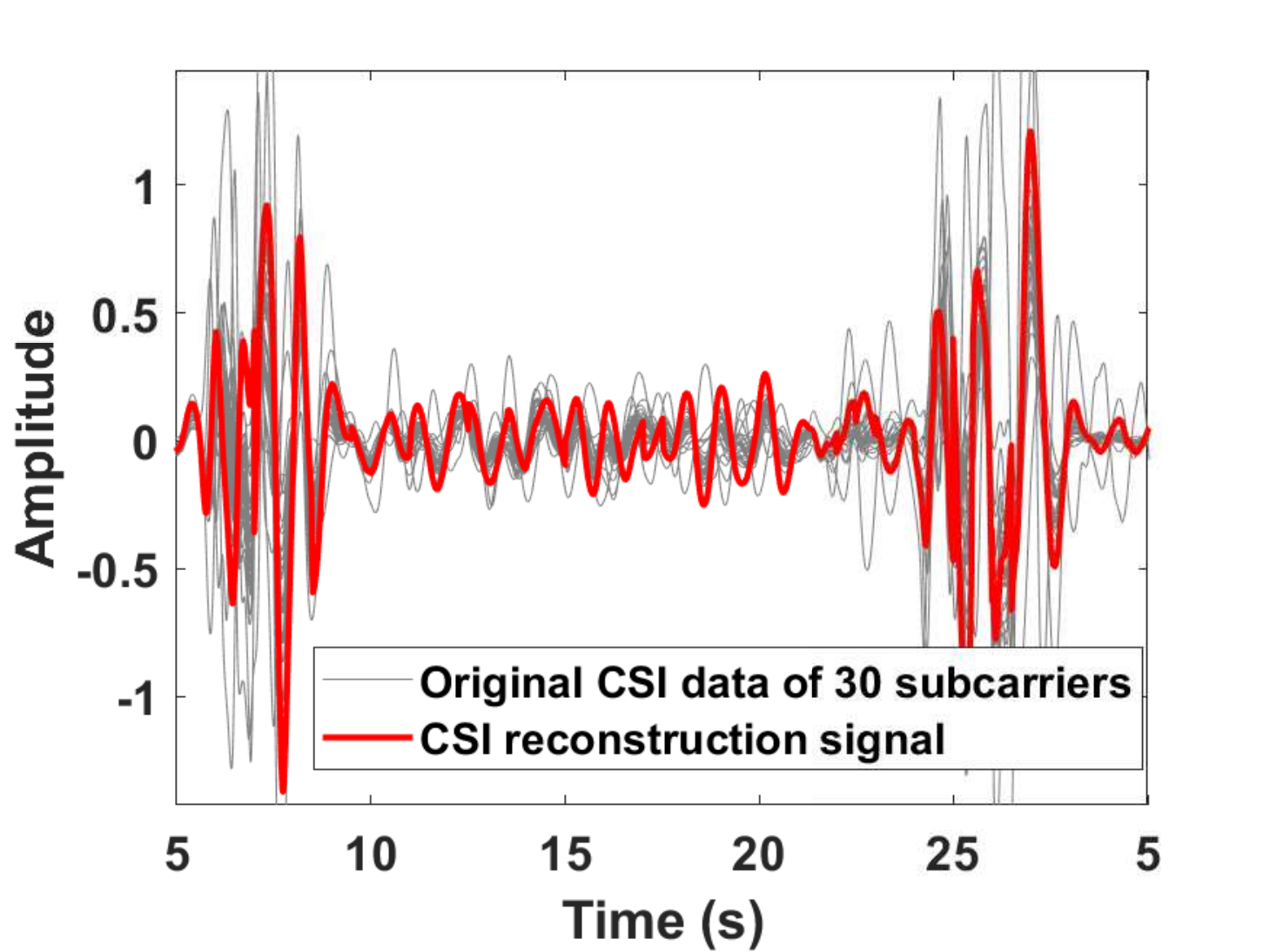}
		\vspace{-6mm}
		\caption{Reconstructing chewing/swallowing on CSI.}
				\vspace{-4mm}
		\label{fig:csi_reconstruction}
	\end{minipage}
	\hfil
	\begin{minipage}{0.45\columnwidth}
		\centering
		\includegraphics[width=1.6in]{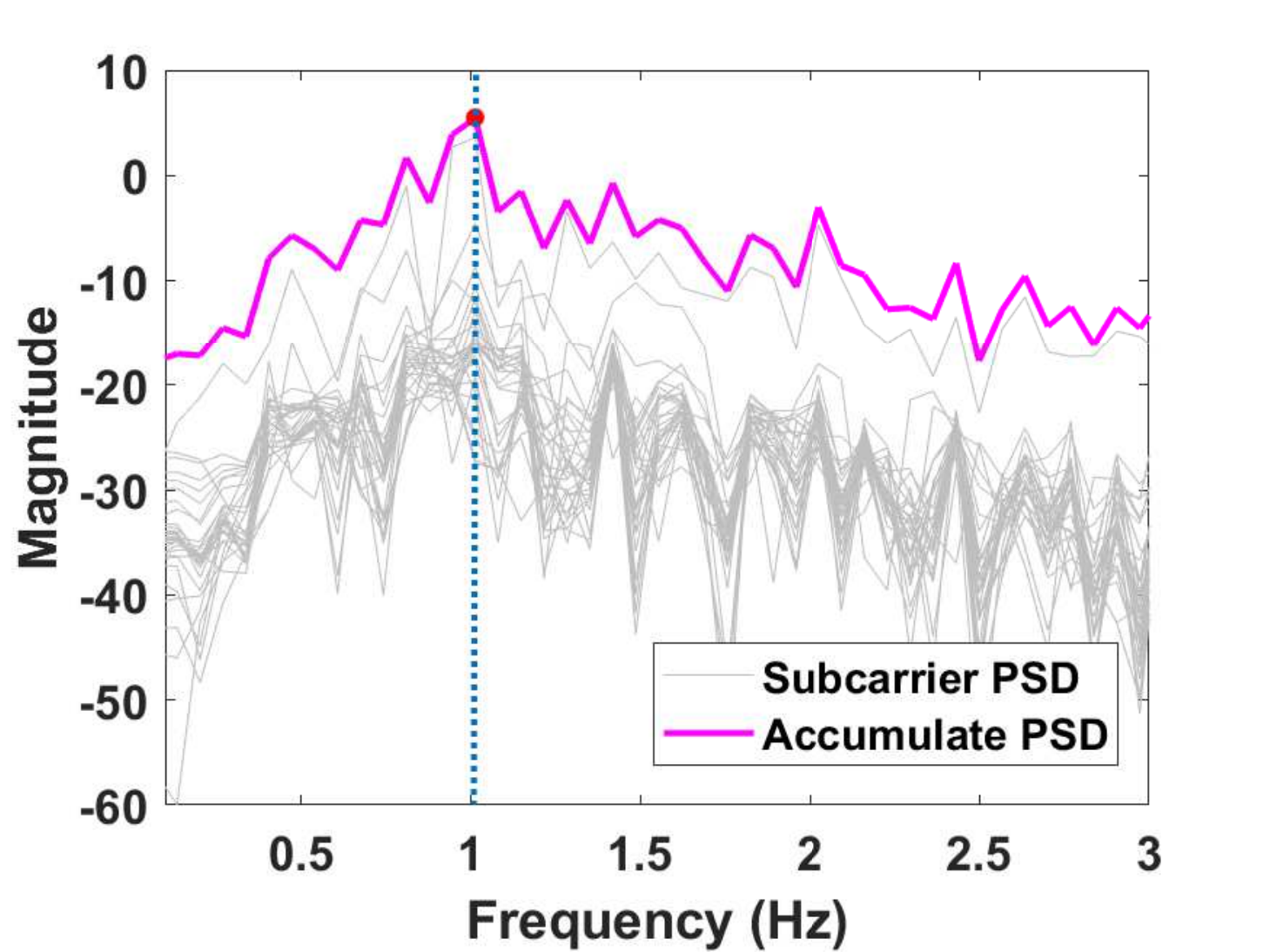}
		\vspace{-6mm}
		\caption{Accumulated PSD of CSI.}
		\vspace{-4mm}
		\label{fig:frequency}
	\end{minipage}
\end{figure}

%

\subsection{Chewing and swallowing Detection}
Chewing is a physical degradation or digestion of food and swallowing is the phase following chewing. To distinguish the food type and calculate the amount of food a person eats, it is essential to detect chewing and swallowing activity.

\subsubsection{Minute Chewing/Swallowing Motion Reconstruction}
Although the chewing and swallowing motions can be seen to have impact (i.e., slight vibrations) on CSI, the minute motion caused by chewing and swallowing still can not be  detected accurately from the CSI. There are two main challenges we need to address.
First, not all the subcarriers are sensitive to tiny motions. Some subcarriers are less sensitive to chewing and swallowing activities because different subcarriers have different central frequencies. Besides the performance of some subcarriers are not consistent, they might present sensitive amplitude for one period time but have dull amplitude for another period time. 
Second, environmental-related uncertainty and noise (i.e., wireless interference and multipath reflection) will also cause CSI fluctuation which makes it hard to differentiate the CSI patterns caused by real chewing and swallowing motions in CSI time series. 

To provide a robust chewing and swallowing detection model, We first utilize a Butter-worth band-pass filter with a cutoff frequency of $0.8$ Hz$\sim$$3$ Hz to remove unwanted noise because chewing and swallowing activities are usually fixed around a low-frequency range~\cite{farooq2016automatic}. Furthermore, a moving average filter is used to remove outliers in the signal.

After we get the filtered signal, based on Month Motion Profile~\cite{wang2016we}, we propose to reconstruct the CSI from all subcarriers to get a single representative CSI sequence, which amplifies the CSI fluctuation caused by chewing and swallowing without losing the original information.
In particular, a sliding window is applied on all subcarriers parallelly and a series of CSI segments are selected from various subcarriers and assembled one by one into a single new CSI according to time series. For each sliding window (the sliding window size is 250ms, which is half the sampling rate), the mean amplitude values of all subcarriers are calculated respectively and then the ten subcarriers which have the nearest mean values to the average mean value of all subcarriers are picked. This step will filter out those abnormal and less-sensitive subcarriers. Furthermore, among these selected subcarriers, the segment which has the maximum peak to peak value within one sliding window is chosen as a representative segment for this time slot. By moving the sliding window along time sequence, we could select a series of CSI segments and produce a single new CSI. As shown in Figure~\ref{fig:csi_reconstruction}, the fluctuation of the new signal is more significant than that from any of the original single subcarrier, which proves the effectiveness of the reconstruction.

\begin{figure}[t]
	\centering
	\includegraphics[width=0.65\columnwidth]{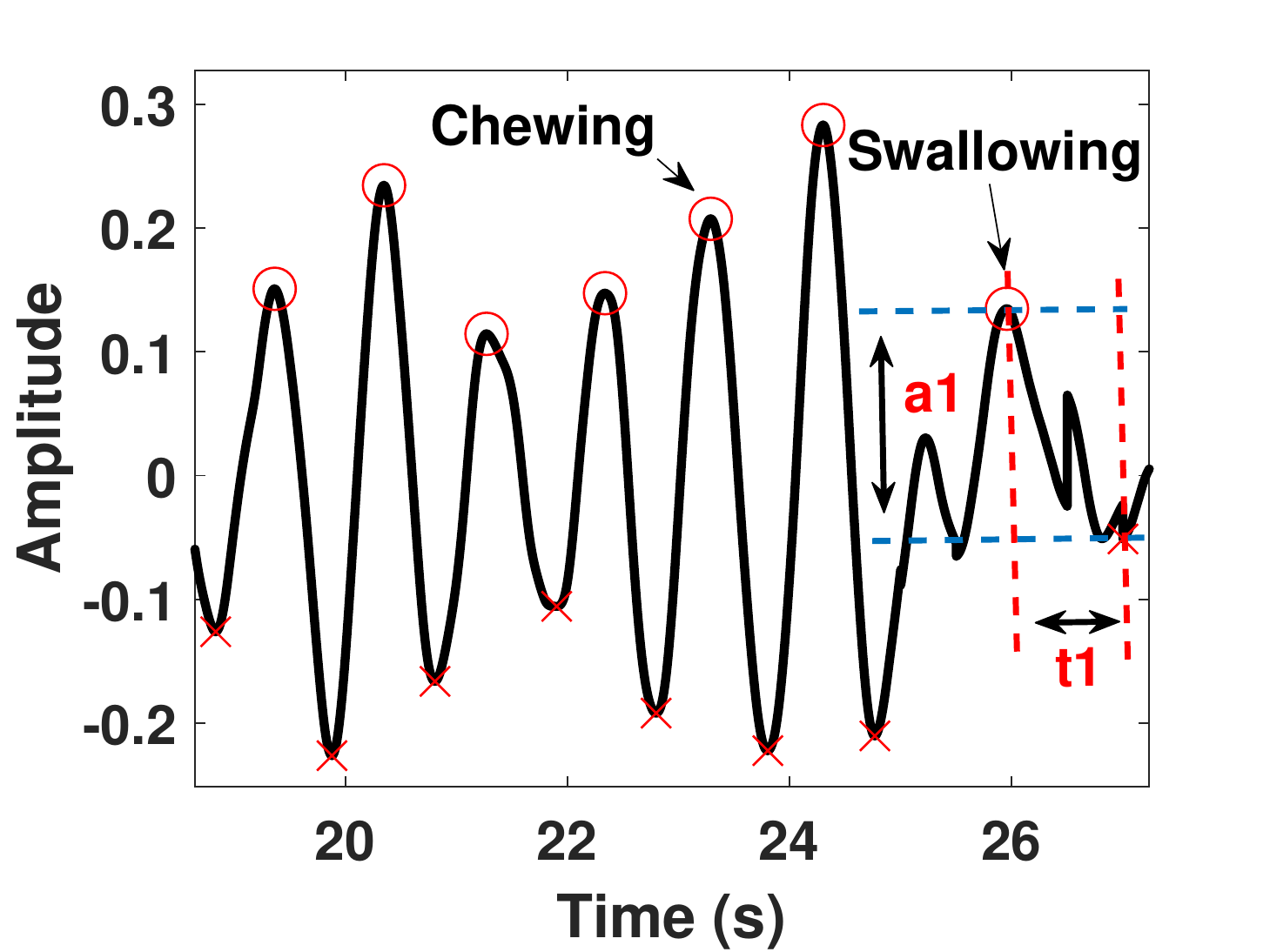}
	\vspace{-3mm}
	\caption{Chewing and swallowing differentiation.}
	\vspace{-5mm}
	\label{fig:classify}
\end{figure}

\subsubsection{Accumulate PSD-based Chewing Period Estimation}
To further reduce the impact of environmental-related noise, and eliminate those fake fluctuation caused by noise, it is necessary to estimate the chewing period.
Based on the estimated period, it is convenient to remove those fake chewing and swallowing motions which are too close to adjacent chewing and swallowing motions.
In our system, instead of calculating the chewing period from time domain, we find it is more effective to acquire a stable period of chewing in frequency domain.
Specifically, we first transfer time domain CSI amplitude to frequency domain by using power spectral density (PSD). 
Since PSD would produce one strong peak at the frequency which corresponding to the periodicity of dominant repetitive fluctuations, we observe that the CSI measurements in our frequency window (i.e., $0.8$ Hz$\sim$$3$ Hz) also present one strong peak which corresponding to the dominant chewing rate.
Some may argue that this strong peak may be dominated by hand movements or environmental factors. However, note that the target CSI measurement is only derived between two adjacent identified eating motions and we have filtered the interferential frequency components by our frequency window. Based on these two reasons, we believe that the estimated period is dominated by mouth motion.

Moreover, to further utilize different subcarriers which present conform results in frequency domain and estimate a more robust chewing period, accumulated power spectral density (APSD) method is proposed in our system to accumulate the results collected from all subcarriers. The accumulated power spectral density of 30 subcarriers with $N$ CSI amplitude measurements can be presented as:

\vspace{-2mm}
\begin{equation}
\footnotesize
APSD = 10log_{10}\sum_{i=1}^{30}\frac{{(abs(FFT(c_{i})))}^2}{N},
\vspace{-2mm}
\end{equation}
where c is the CSI measurements of subcarrier $i$.
Since the swallowing times is far less than chewing times in real life, the highest peak is identified as the chewing rate of the specific period between two eating motions. Figure~\ref{fig:frequency} shows an example to infer the chewing rate of one period time between two eating motions. The ground truth of the chewing rate is $1$Hz and Figure~\ref{fig:frequency} depicts that there is one strong peak around 1HZ in the APSD which implies that our APSD could effectively estimate the chewing rate of a user.

\begin{figure}[t]
	\centering
	\subfigure[Smartphone-AP Setup]
	{\includegraphics[width=.48\columnwidth]{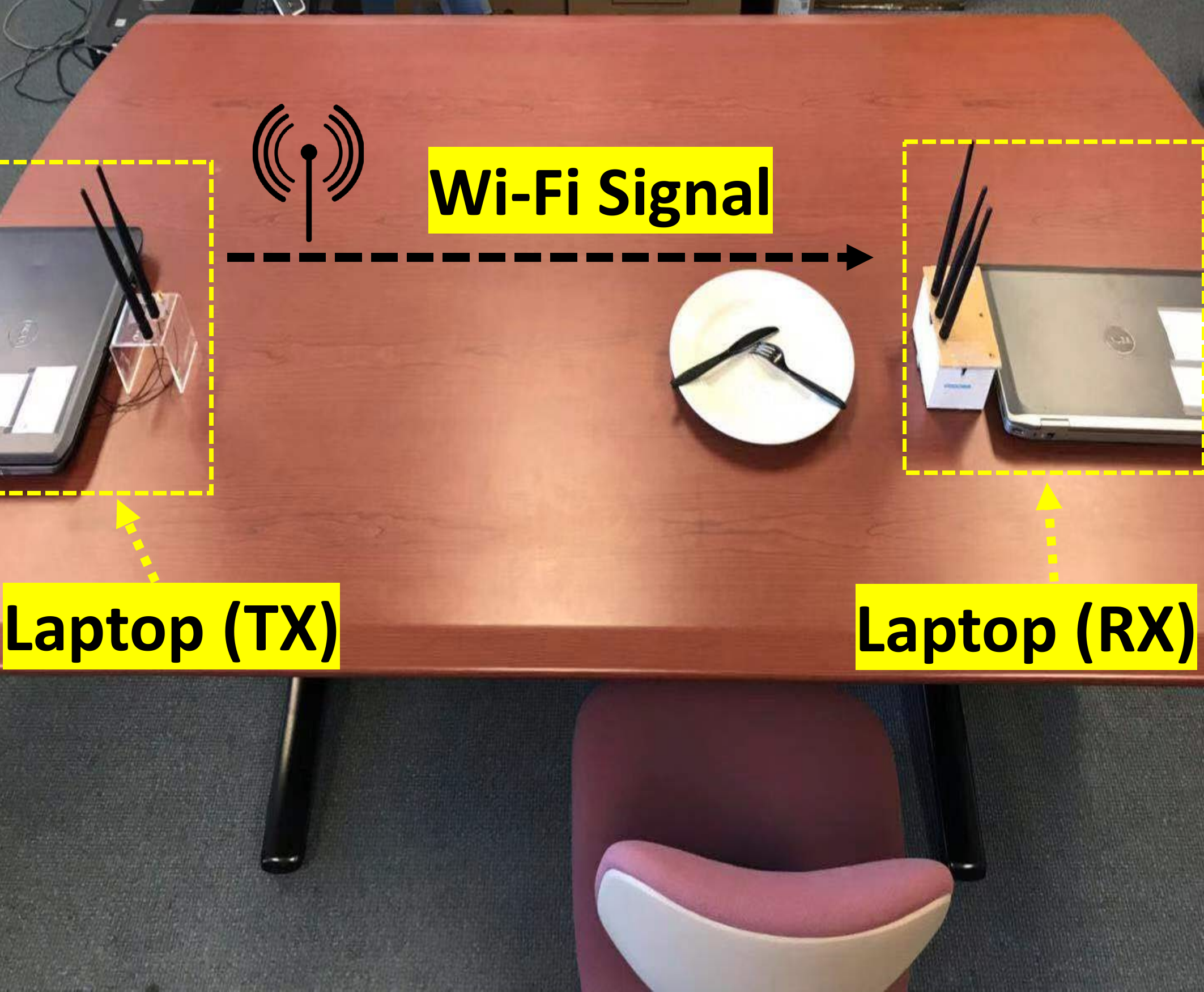}\label{fig:experiment_setup}}
	\subfigure[IoT device-AP Setup]
	{\includegraphics[width=.48\columnwidth]{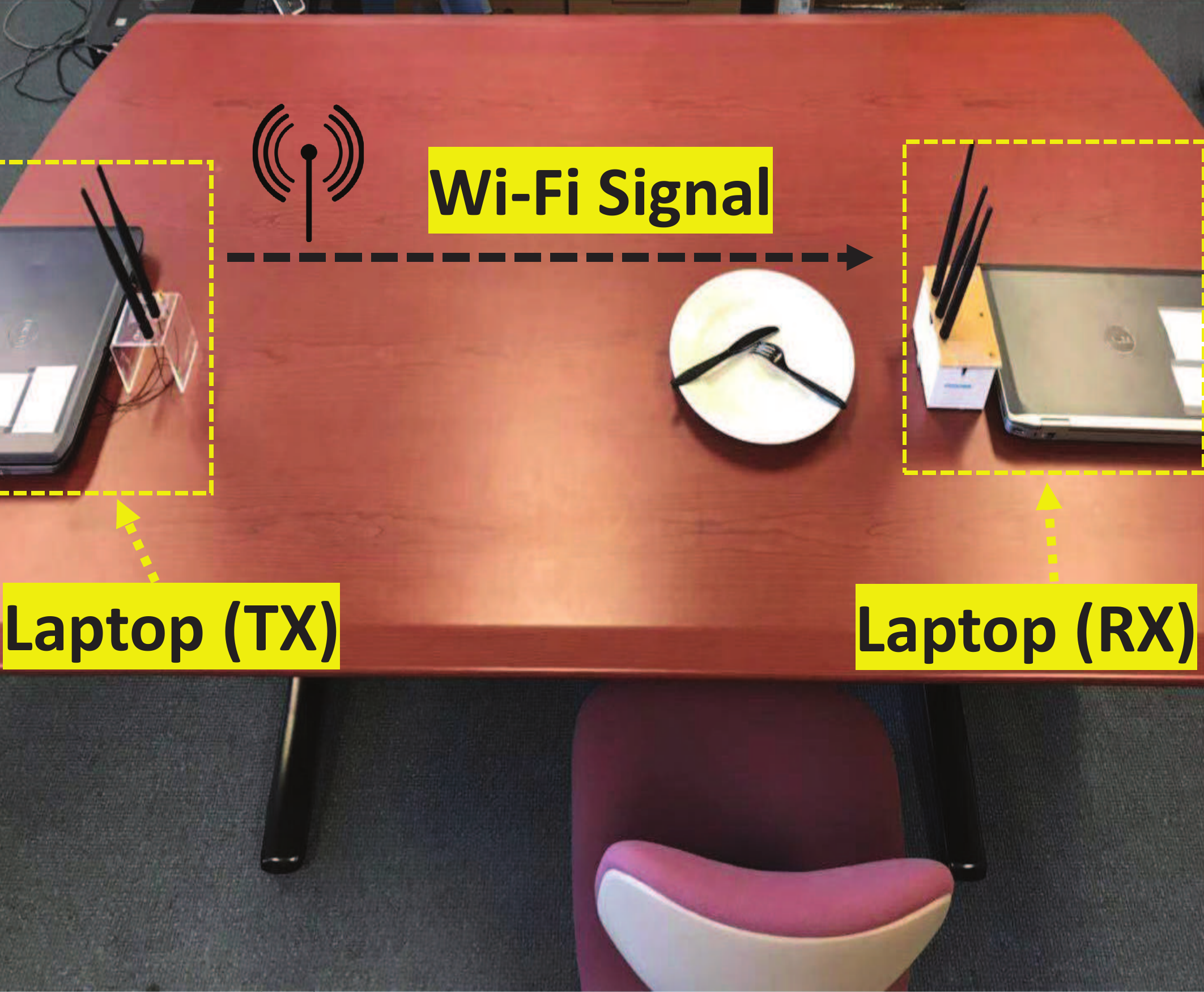}\label{fig:experiment_setup_iot}}
	\vspace{-2mm}
	\caption{Illustrations of two experimental settings.}
	\label{fig:experiment_settings}
	\vspace{-6mm}
\end{figure}

\begin{figure*}[t]
	\centering
	\subfigure[Eating moments recognition]
	{\includegraphics[width=0.25\textwidth, clip=true]{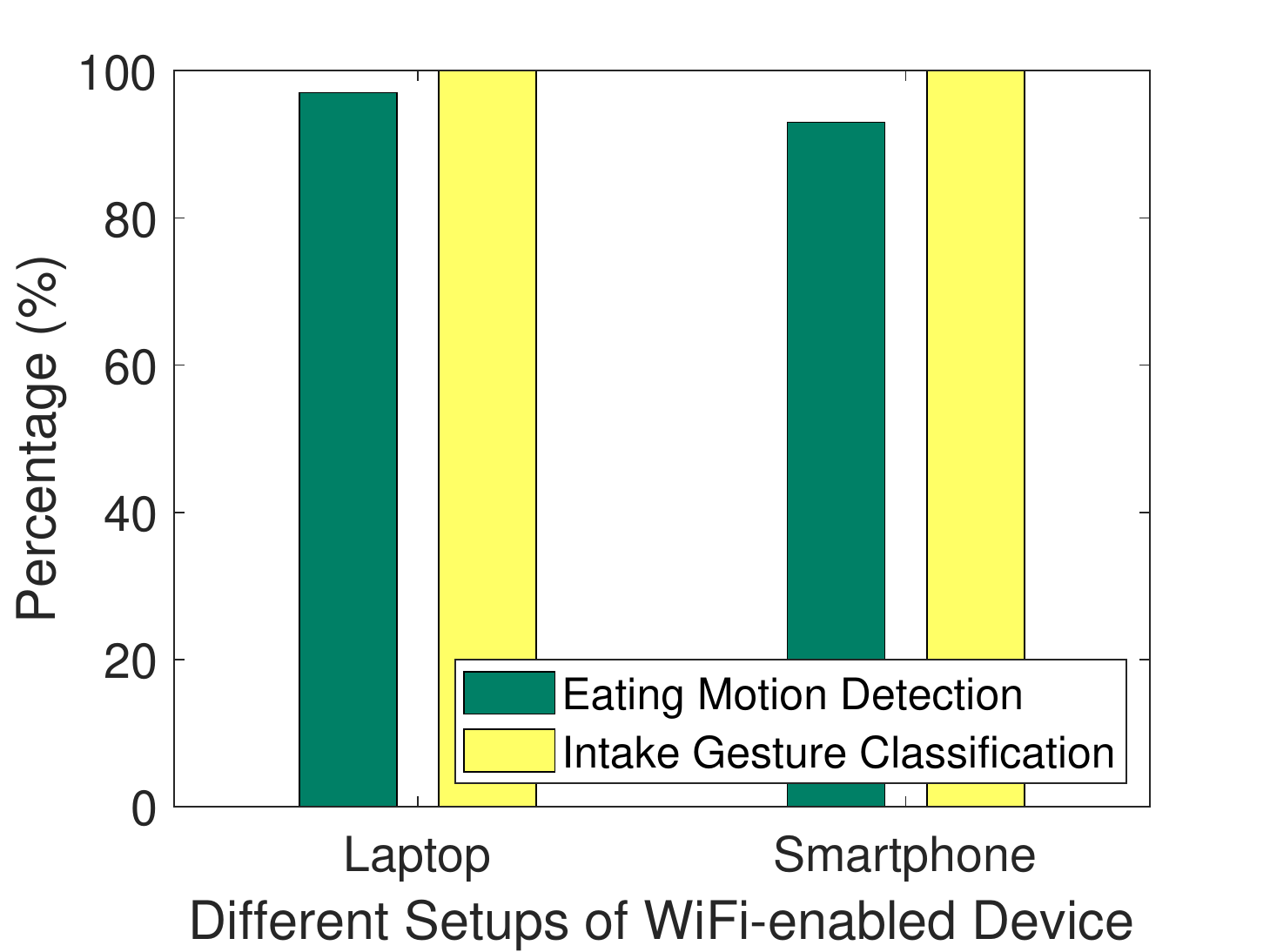}\label{fig:eating and non-eating setups}}
	\subfigure[Utensils identification]
	{\includegraphics[width=0.25\textwidth, clip=true]{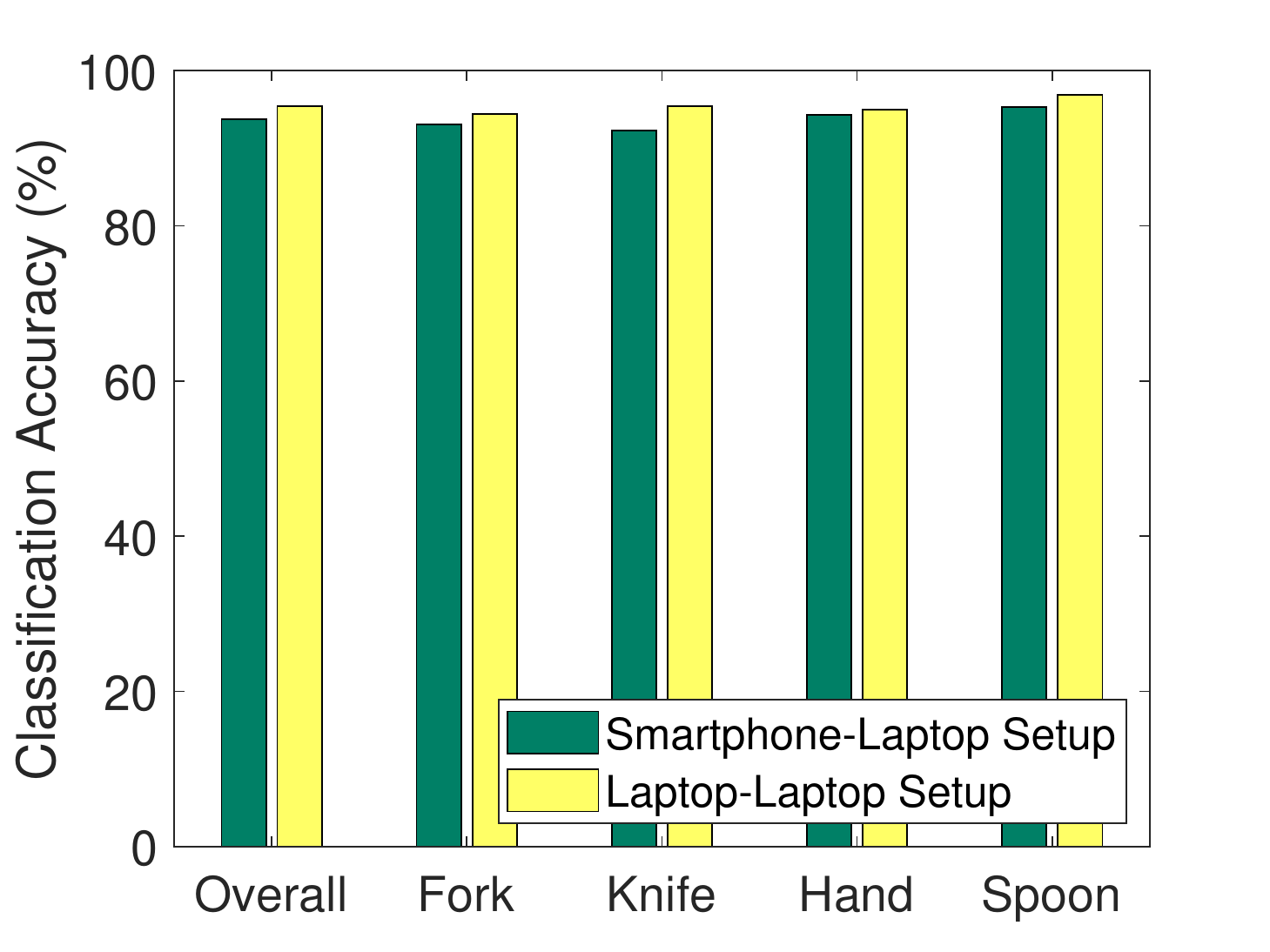}\label{fig:eating motions setups}}
	\subfigure[Chewing and swallowing detection]
	{\includegraphics[width=0.25\textwidth, clip=true]{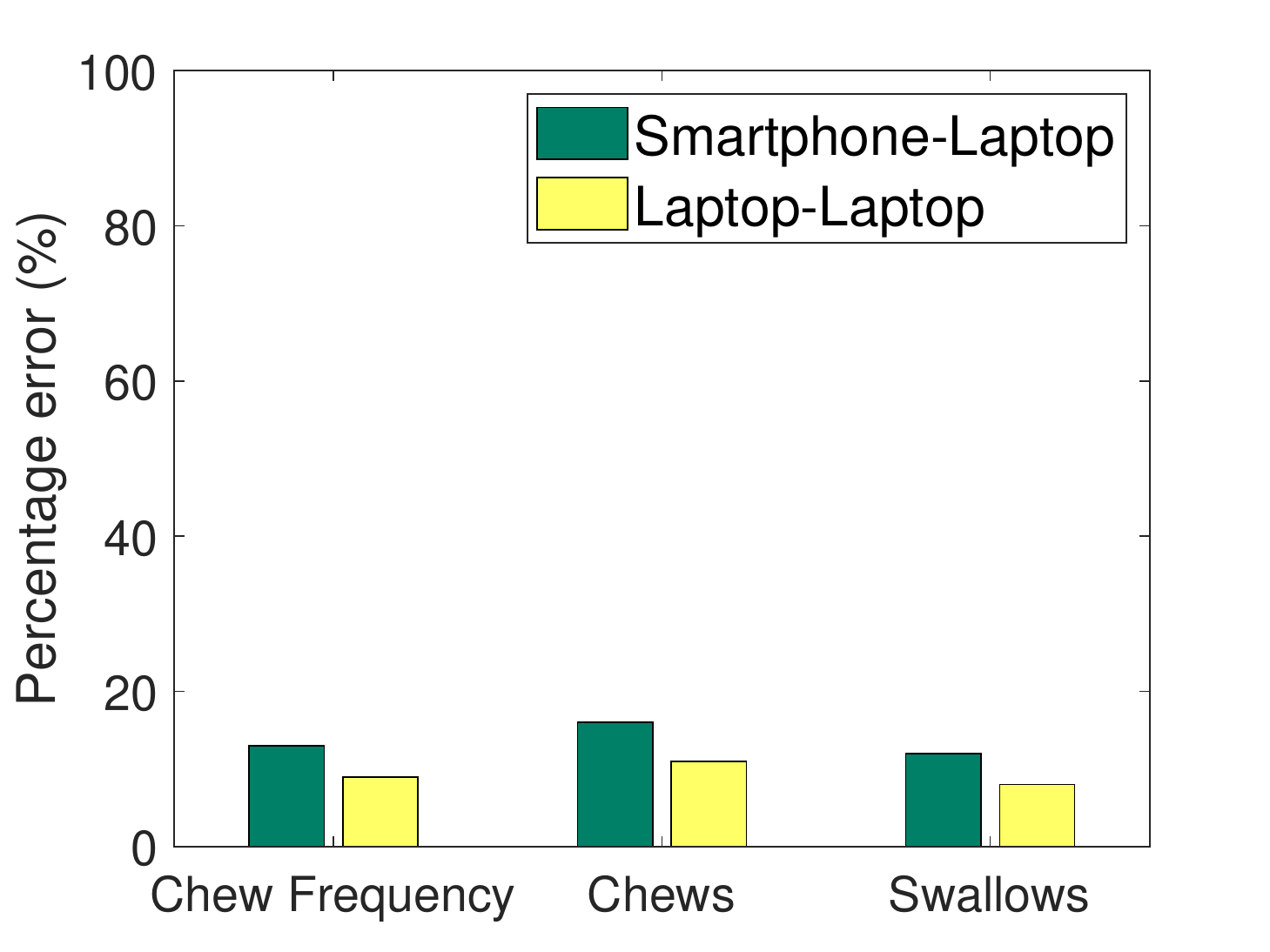}\label{fig:chewing and swallowing setups}}
	\vspace{-2mm}
	\caption{Performance comparison of WiEat under the smartphone-AP setup and the IoT device-AP setup.}
	\label{fig:setup}
	\vspace{-5mm}
\end{figure*}

Based on the average period of chewing, we propose a threshold-based peak detection approach to identify the candidate CSI patterns caused by chewing and swallowing on CSI data.
In original peak detection algorithm, a peak could be found if the value of data is larger than its two neighboring data.
However, this algorithm has two limitations. 
First of all, some peaks which are too close to each other that are normally caused by one chewing and swallowing motion can all be marked out.
Second, some peaks caused by environmental noises could also produce some tiny peaks.
In our system, two thresholds are set empirically to remove fake peaks.
Firstly, based on the average chewing period, a minimal distance $\beta$ between two neighbor peaks is used to restrict the peak-to-peak separation and only peaks that recur at regular intervals are preserved.
Second, a minimum amplitude of the peaks $\gamma$ is used to filter out those fake peaks caused by environmental noises. 
After filtering these fake peaks, we get the number of all CSI patterns caused by mouth motions, and in the next part, we can distinguish chewing swallowing motions from those preserved real peaks.

\subsubsection{Threshold-based Swallowing Detection}

Swallowing counting is challenging because the CSI measurements between chewing and swallowing are both sinusoidal-like patterns. Nevertheless, we observe that the chewing and swallowing of CSI measurements are still distinguishable on the reconstructed CSI representative. Compared with chewing motions, swallowing motions are more slowly and the movement range of the throat when swallowing is more moderate than jaw activity.
We therefore propose a threshold-based mechanism to classify the CSI patterns into two classes based on two measurements, including CSI amplitude range and corresponding peak-to-valley time interval.
Specifically, as shown in Figure~\ref{fig:classify}, a swallowing motion occurs following several times of chewing. The peak-to-valley time interval of CSI sinusoidal patterns is calculated as $t_{j} = l_{j+1}-l_{j}$, while the range of CSI amplitude between the peak and corresponding valley is derived from $A_{j} = p_{j}-v_{j}$. $p_{j}$ and $v_{j}$ are corresponding to the peak and valley of the $j^{th}$ CSI sinusoidal pattern respectively, and $l_{j+1}$ and $l{j}$ are the time stamp of the peak and valley of the same sinusoidal pattern on time serials. 
After we obtain the measurements of each CSI sinusoidal pattern. A threshold-based method is proposed to classify the candidate CSI sinusoidal pattern into two classes. The average values of the amplitude $\overline{a}$ and period $\overline{t}$ of all CSI sinusoidal patterns within a specific period time (e.g. between two adjacent eating motions.) are identified as the reference threshold of these two classes. Those CSI sinusoidal patterns whose amplitude $a_{j}$ is less than $\overline{a}$ and also the period $t_{j}$ is larger than $\overline{t}$ are counted as $sw_{i}$ which represents one swallowing time. Once we recognize swallowing from chewing, it is very convenient to derive the statistic of chewing and swallowing respectively.

%% file: text_infocom/performance.tex
\section{Performance Evaluation}

\subsection{Experimental Methodology}

\textbf{Experimental Setup.}
We implement our system with a pair of WiFi devices in two setups as shown in Figure~\ref{fig:experiment_settings}. The \textit{Smartphone-AP Setup} describe the practical scenario when the user places the personal smartphone on the dinning table during eating. In this setup, the smartphone is connected to a WiFi AP to receive WiFi signals and sense the user's eating activities. \textit{IoT Device-AP Setup} describe a different scenario, where an IoT device (e.g., smart TVs and restaurant table tablets ) can use the WiFi signals received from an AP to capture the user's eating. Both setups are deployed in three representative indoor environments to evaluate our system, including a small office, a large office and a dining room. 

\textbf{Devices. }We conduct experiments with two different smartphone models, Nexus $6$ and Huawei Mate $10$ samrtphones. The Nexus $6$ has 3 GB RAM and a 2.7 GHz Snapdragon 805 processor while the Huawei Mate $10$ is equipped with a 4GB RAM and a 2.36 GHz Kirin 970 processor. We also utilize two Dell E$6430$ laptops equipped with 802.11n WiFi wireless card IWL 5300 NICs~\cite{halperin2011tool} and 6dBi rubber ducky external omni-directional antennas for extracting CSI data. One laptop is used to imitate the IoT device and the other serves as the AP. Both laptops are running Ubuntu 10.04 LTS with the kernel 2.6.36. And the WiFi cards work at 5GHz frequency band with transmission rate $500 pkt/sec$. 

\textbf{Data Collection. }
We recruit $20$ participants to perform eating activities with the two experimental setups at three representative indoor environments. In total, $1600$ minute eating period data is collected. 
To collect CSI in the smartphone-AP setup, we configure the laptop (serving as an AP) to run in the netlink mode, which sends Internet Control Message Protocol (ICMP) echo and gets the reply from the smartphone to collect the CSI data from the smartphone~\cite{li2016csi}. For the IoT device-AP setup, we configure the both laptops (i.e., an IoT device and an AP) to work under the injection mode and the IoT device directly collect the CSI data. Moreover, for both setups, we test three distances ($80$cm, $100$cm, and $120$cm)and four different placements to deploy the devices. 

\textbf{Evaluation Metrics.}
$Detection \;  Rate$ is the ratio of the number of correctly detected eating activities over the total number of eating activities.
$Accuracy$ is the ratio of the number of correctly detected activities over the total number of activities.
$Percentage \; Error$ of estimating the chewing/swallowing count is defined as: 

$percentage \; error = \frac{\left |  estimated \; number - ground \; truth \right |}{ground \; truth}$.


%
%
%

\subsection{Performance of Smartphone-AP and IoT Device-AP Setups}

We first evaluate the performance of our eating monitoring system in the smartphone-AP setup and then compare its performance with the IoT device-AP setup, when the AP is placed 80 cm at the left side of the smartphone/IoT device.  


\subsubsection{Differentiate eating and non-eating activities}
Figure~\ref{fig:eating and non-eating setups} presents the accuracy and detection rate of our proposed system when smartphones and IoT devices are used.
We can observe that the accuracy for the eating/non-eating activity detection is over 87\% on smartphones and 92\% on IoT devices. This is because the laptops have better reception of WiFi signals with the external antenna. Whereas in the eating activity recognition, we find that our system could achieve a detection rate of 100\% on both smartphones and IoT devices and it also confirms that the effectiveness and reliability of our system on eating activity detection.

\subsubsection{Identify eating motions with utensils}
Figure~\ref{fig:eating motions setups} depicts the performance of our soft decision-based classification algorithm on differentiating eating utensils including spoon, fork, fork \& knife and hands.
We observe that our system can achieve over 90\% accuracy across different eating utensils for both smartphone and IoT device setup.
Besides, it is encouraging to find that all the four types of utensils can all be recognized well with the lowest accuracy as $90.05\%$ on smartphones and $93.3\%$ on IoT devices.
This demonstrates that our system is robust for utensils identification even if the different device setups are used.

\subsubsection{Chewing and swallowing activities detection}
Figure~\ref{fig:chewing and swallowing setups}  illustrates the performance of our system on chewing and swallowing detection.
This figure shows that the percentage error of our system on chewing frequency is below $15$\% on both smartphones (i.e., 13\%) and IoT devices (i.e., 9\%).
Further, we also observe that the percentage error for chewing number estimation and swallowing number estimation are all below $16$ \% under all scenarios.
These results confirm the effectiveness of our system on  behavior recognition under various eating stages.

\begin{figure}[t]
	\centering
	\subfigure[Differentiate eating/non-eating]{\includegraphics[width=1.6in]{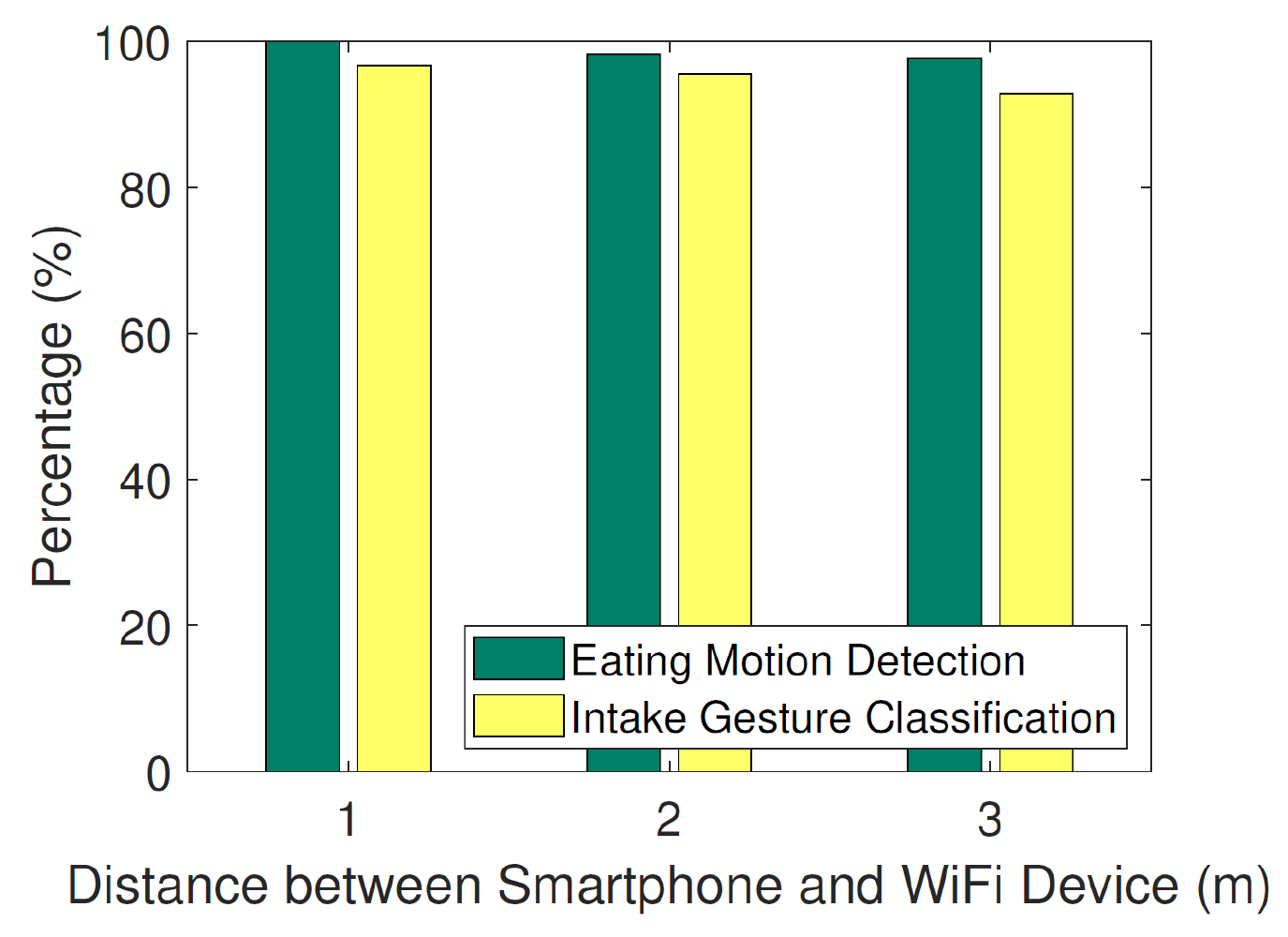}\label{fig:eating and non-eating distance}}
	\subfigure[Identify detailed eating motions]{\includegraphics[width=1.6in]{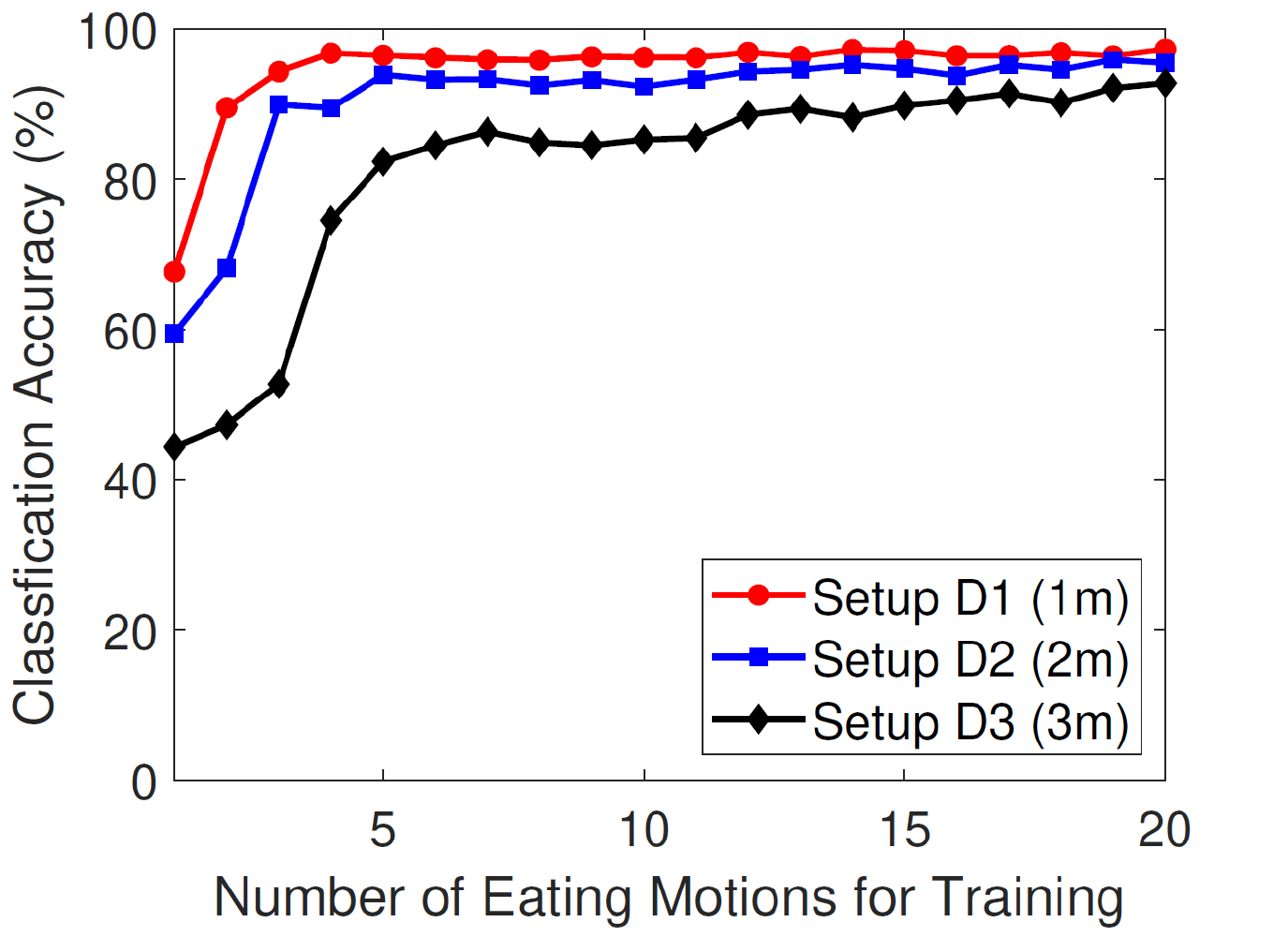}\label{fig:eating motions distance}}
	\subfigure[Chewing detection]{\includegraphics[width=1.6in]{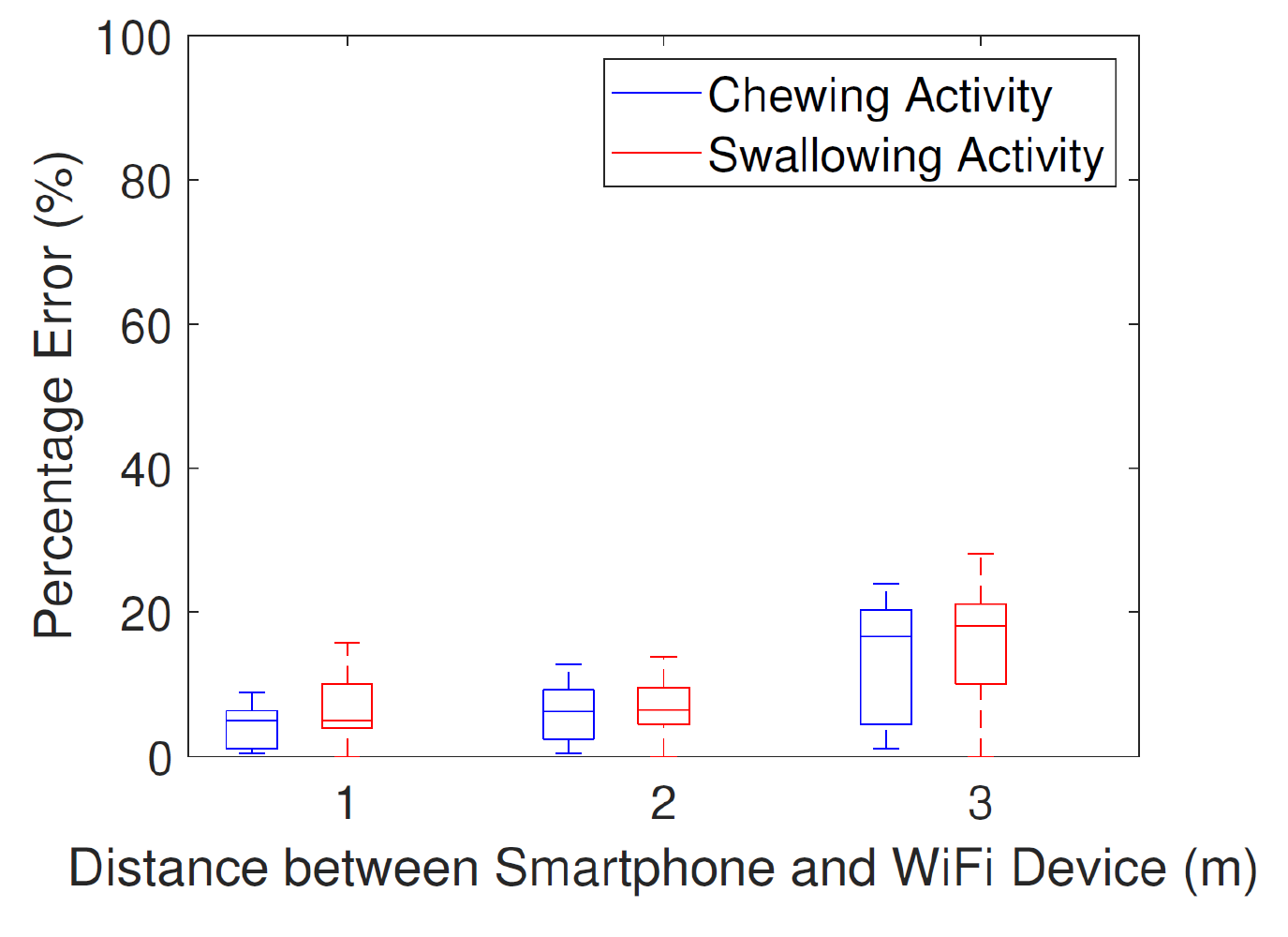}\label{fig:chewing distance}}
	\subfigure[Swallowing detection]{\includegraphics[width=1.6in]{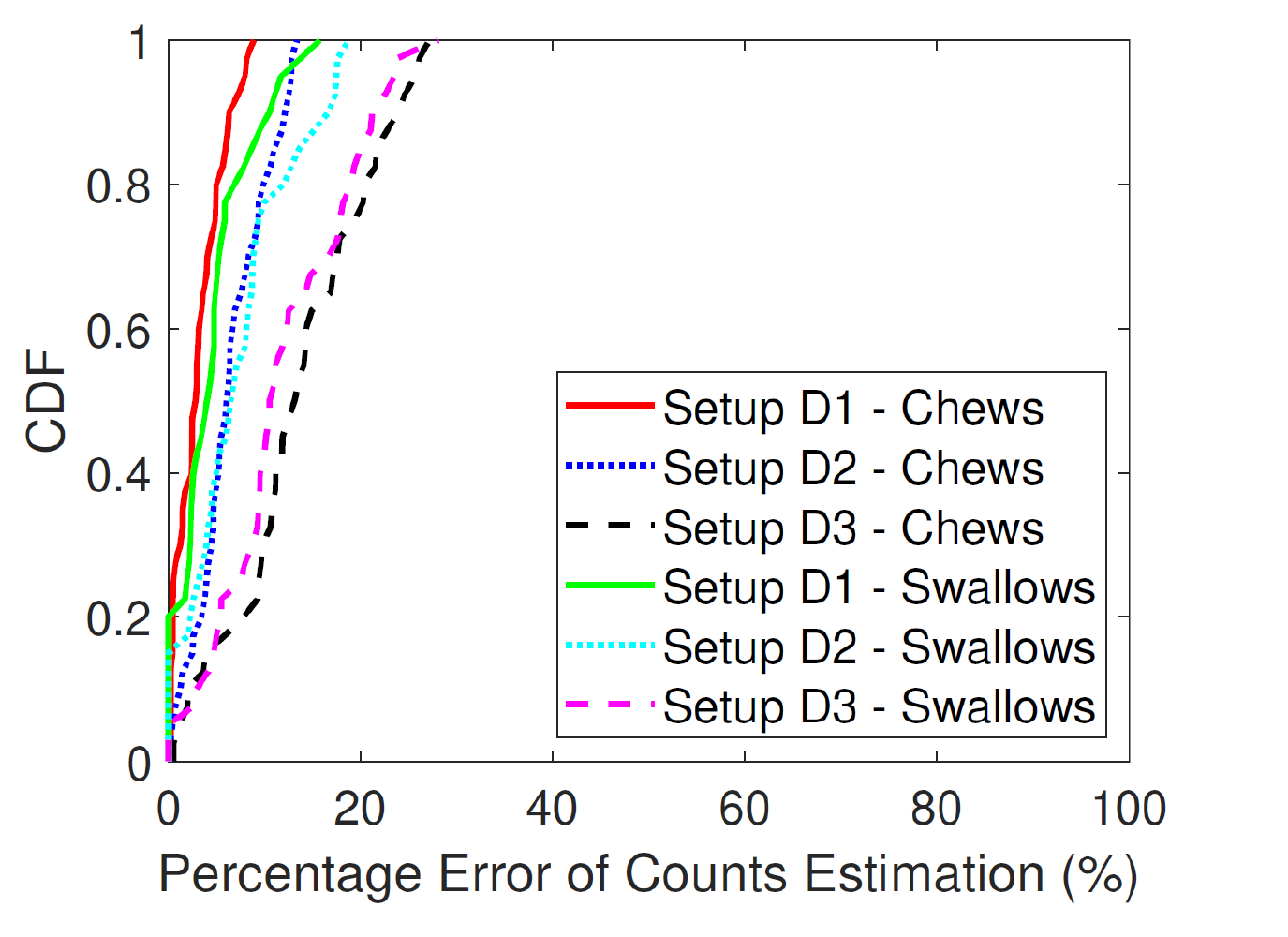}\label{fig:swallowing distance}}
	\vspace{-2mm}
	\caption{Impact of the AP's distance to the performance of WiEat.}
	\vspace{-5mm}
	\label{fig:distance}
\end{figure}

\subsection{Impact of AP Distance}

According to the common dietary customs of participants in real dining environments, we assume that the smartphone is always placed on the right-side of participants in a dining table. Since the length of the table is limited, we choose a large dinging room to explore the performance of eating motions, chewing and swallowing rate estimation under various distances. The AP and the smartphone device are placed at two sides of the dining table with distances under 80cm, 100cm, 120cm, respectively. Figure~\ref{fig:eating and non-eating distance} presents the detection rate and accuracy in terms of eating activities recognition under different distances. Overall, we observe that our algorithm can achieve over 96\% detection rate and 91\% accuracy under different distances, which demonstrates that our system is very accurate across different distances including longer distances such as 120cm when participants sat far away from the AP. Figure~\ref{fig:eating motions distance} depicts the overall identification accuracy of eating motions with utensils is higher than 85\% under different distance. As for chewing and swallowing detection, we compare the results in different distances, finding that under 120cm distance, the percentage error of these minute motions derivation will slightly increase. This is because the minute mouth and throat movements involved in the ingestion process are hard to be captured due to the weak strength of WiFi signal in longer distance. In addition, shorter distance between the AP and the smartphone device lead to better performance in terms of the received WiFi signals are stronger with shorter communication distances. Figure~\ref{fig:chewing distance} and Figure~\ref{fig:swallowing distance} showing that, even for the 120cm situation, our system can achieve the mean percentage error of the chewing and swallowing detection are below than 0.35. The results show that our system provides accurate eating monitoring when the AP is placed at different distances away. Moreover, since we are increasingly addicted to mobile devices nowadays, it is very common for people to keep their phones in sight and glue to phone screen while eating, which actually provide the practicality of our system that using a smartphone to monitor eating activities in daily life. We also believe that with the popularity of Internet of things in recent years, more and more surrounding IoT devices (e.g., smart TVs, smart table, smart light) can be utilized as the WiFi signal receiver to capture the users' eating data , which might further increase the potentiality and performance of our method.

\begin{figure}[t]
	\centering
	\subfigure[Differentiate eating/non-eating]{\includegraphics[width=1.6in]{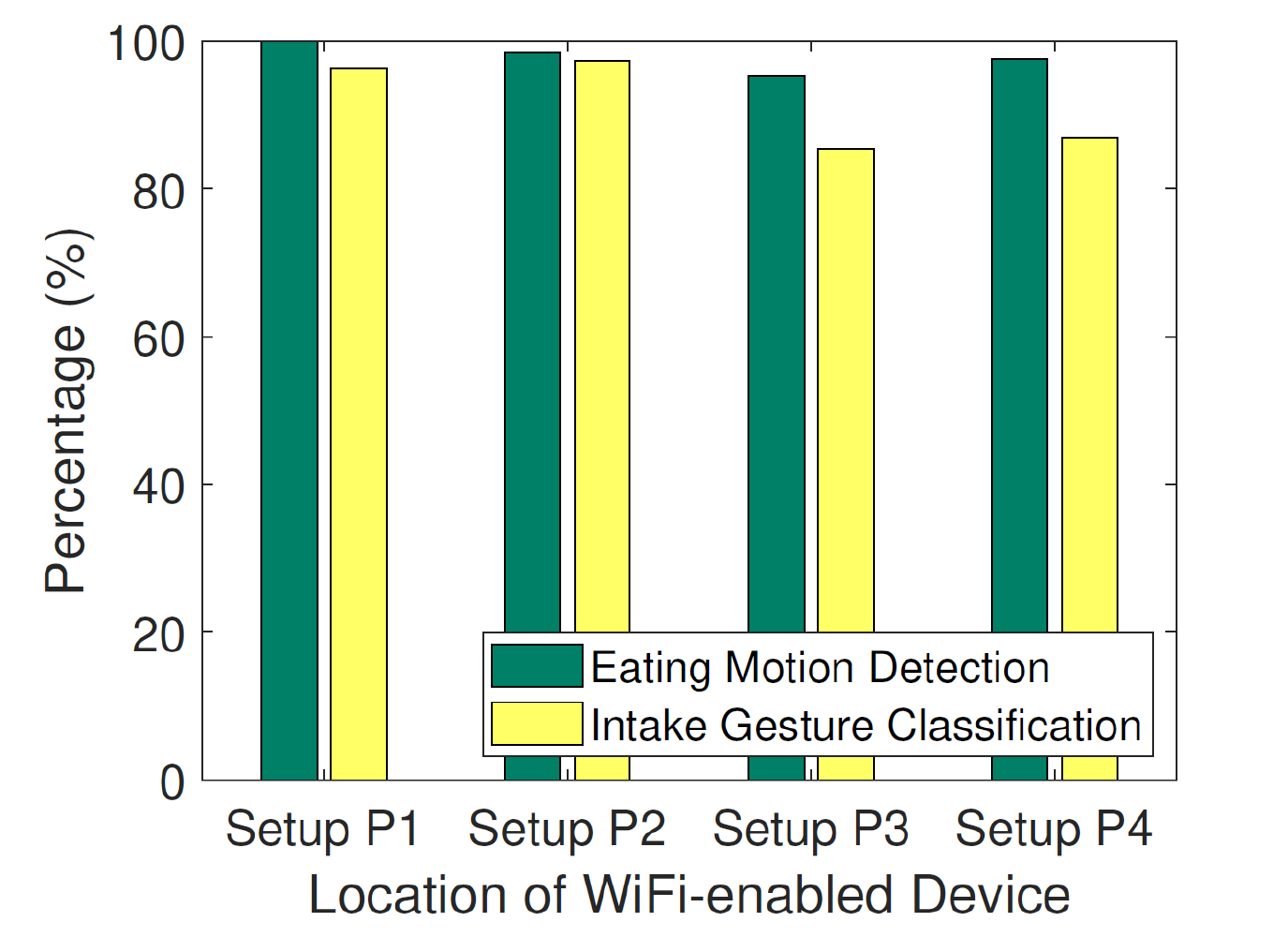}\label{fig:eating and non-eating displacements}}
	\subfigure[Identify detailed eating motions]{\includegraphics[width=1.6in]{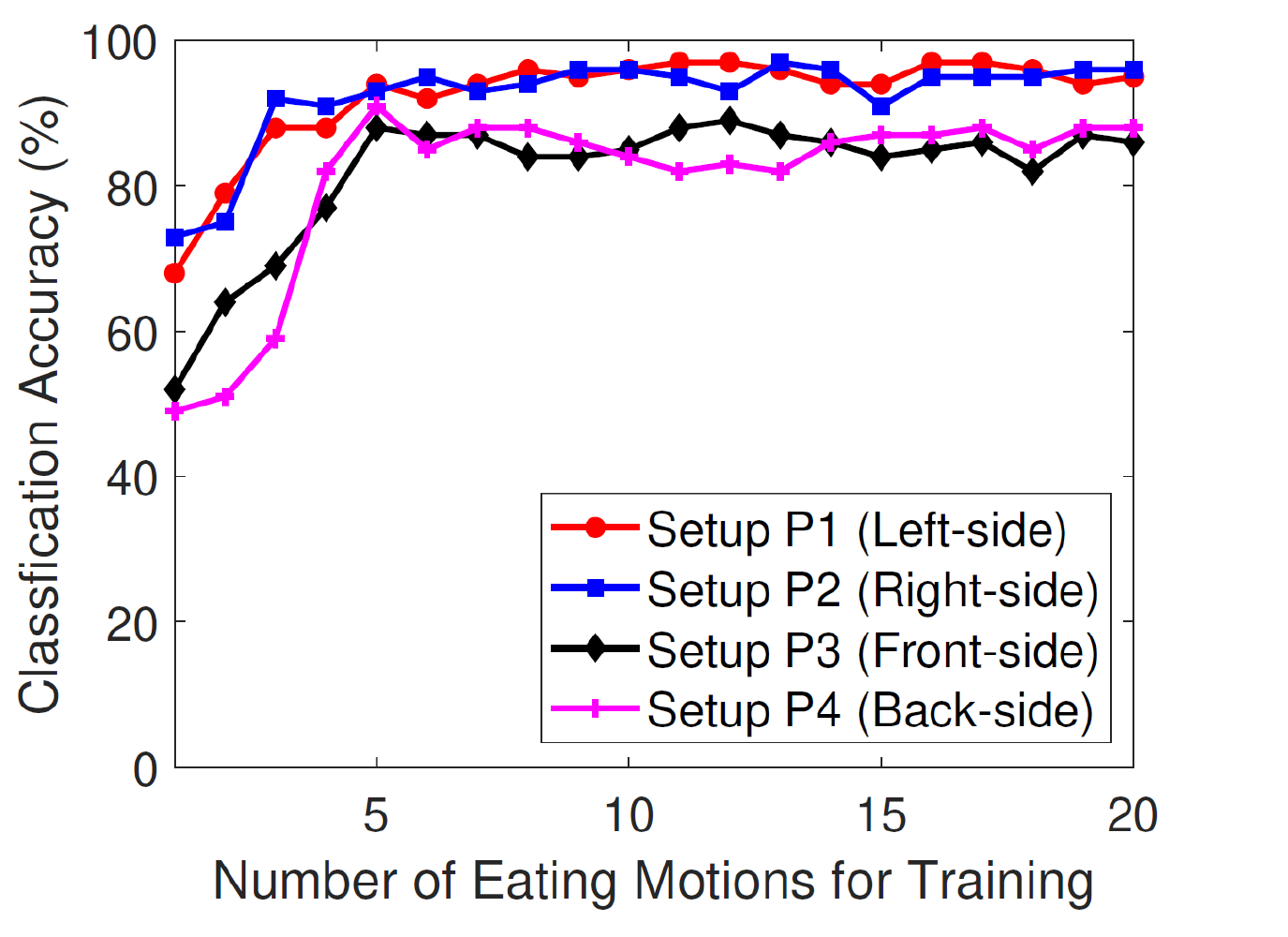}\label{fig:eating motions displacements}}
	\subfigure[Chewing detection]{\includegraphics[width=1.6in]{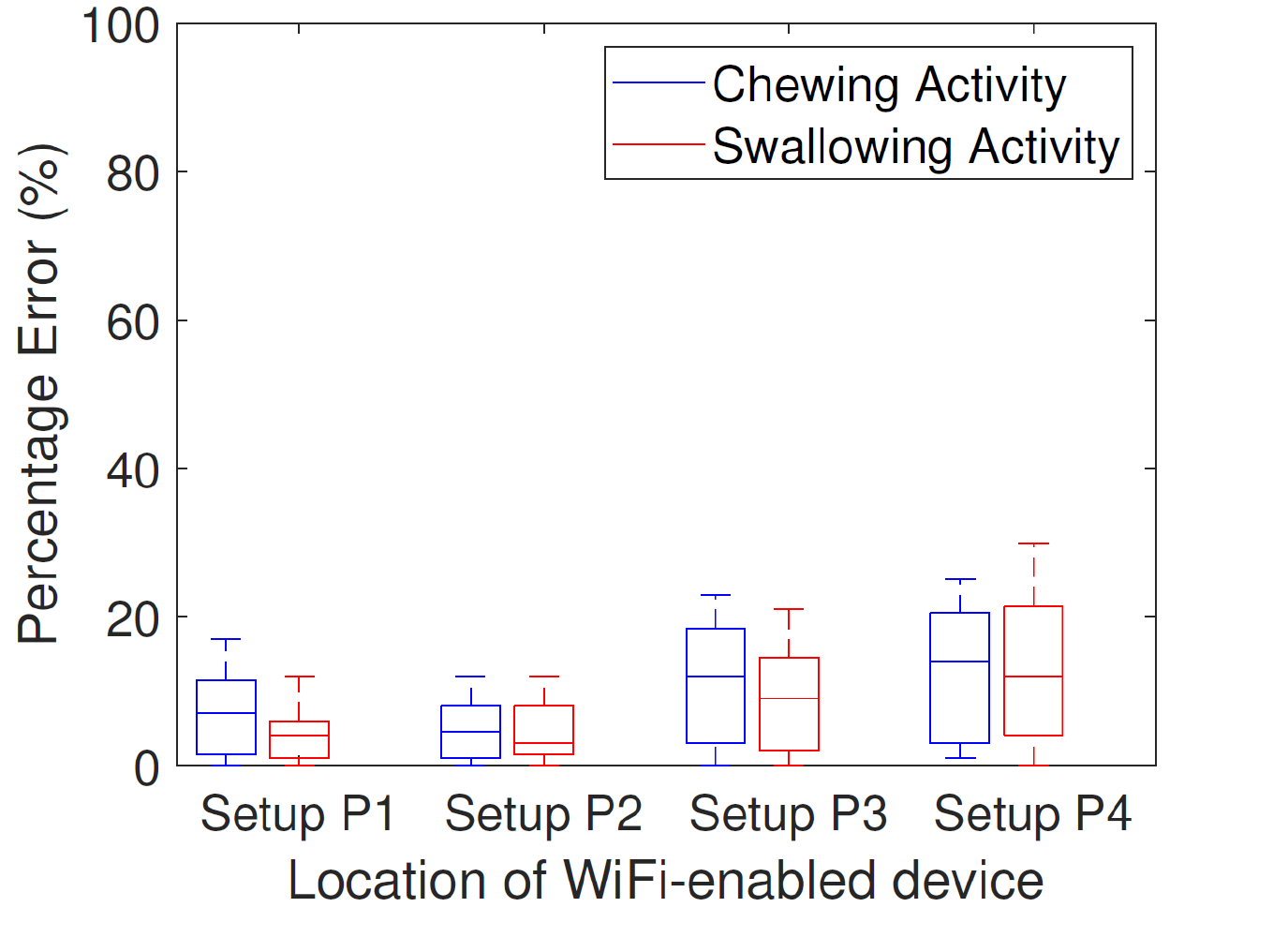}\label{fig:chewing displacements}}
	\subfigure[Swallowing detection]{\includegraphics[width=1.6in]{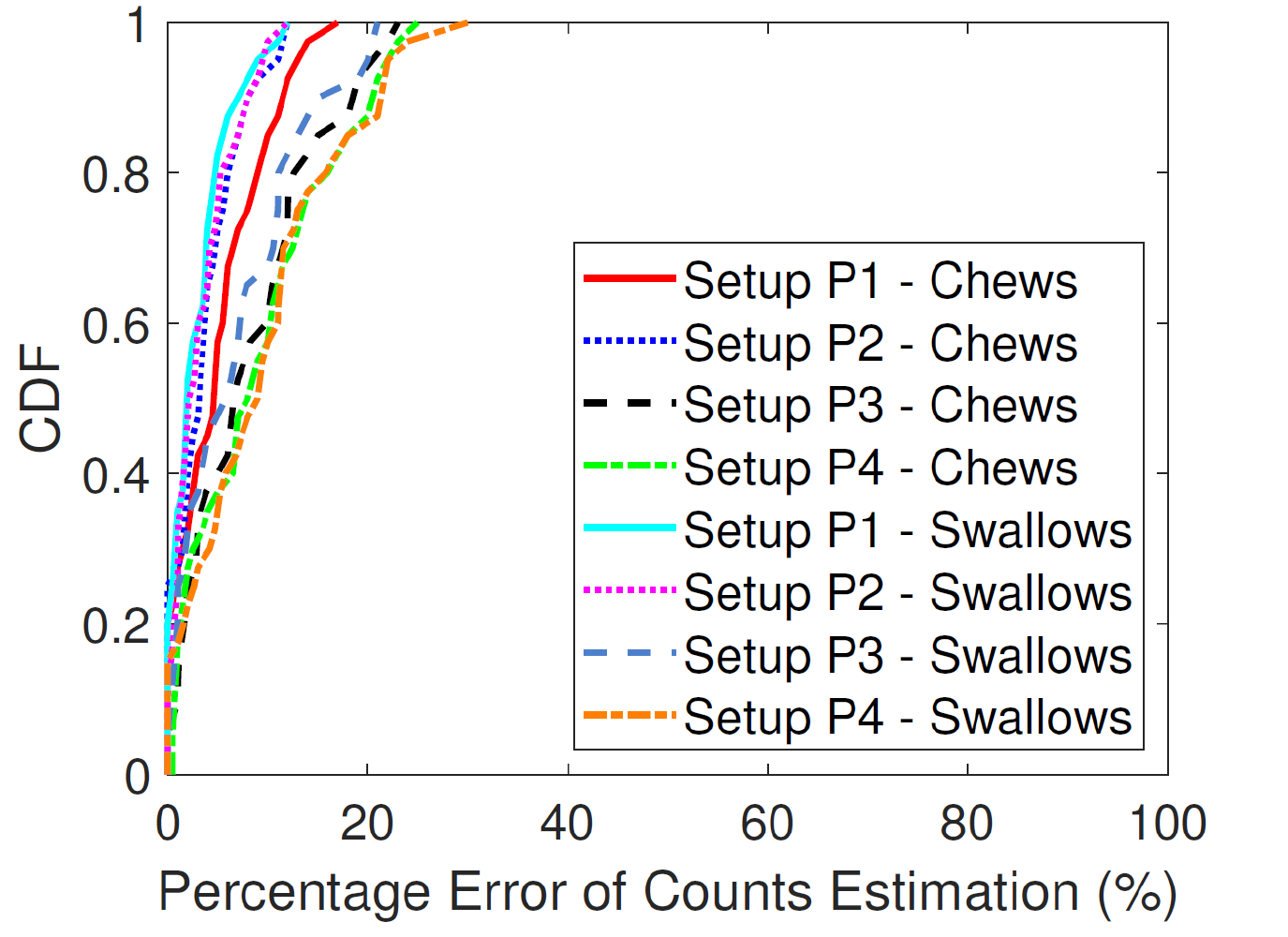}\label{fig:swallowing displacements}}
	\vspace{-2mm}
	\caption{Impact of the AP's displacement to the performance of WiEat.}
	\vspace{-5mm}
	\label{displacement}
\end{figure}

\subsection{Impact of AP Displacement}

Figure~\ref{fig:eating and non-eating displacements} shows the accuracy and detection rate of differentiate eating from non-eating under different placements of the WiFi AP when the smartphone is close to the user, as shown in Figure~\ref{displacement}. We find that the eating activities recognition has the higher accuracy when placing the AP at the left-side (96\%) or right-side (92\%). In addition, Figure~\ref{fig:eating motions displacements} depicts the accuracy of identifying the detailed eating motions with utensils. We observe that the left-side and the right-side settings show better performance than the front-side and back-side settings. The major reason is that the human body blocks the LOS in the front-side and back-side settings, which dominates the interferences of the WiFi signals causing the hand motions and jaw movements harder to be detected. Figure~\ref{fig:chewing displacements} and Figure~\ref{fig:swallowing displacements} show that the mean percentage error of the chewing and swallowing detection are below than 0.4. The above results show that our system is effective under different relative displacements of the AP but achieve the best performance under the left-side and right-side settings.

%% file: text_infocom/discussion.tex
\section{Discussion}



\subsection{Personalized Monitoring System}
According to the motivation of our system, our system aims to provide a personalized machine learning-based eating monitoring method; thus we collect the users' personal training data and built eating profile under a specific setting and then test the performance of our system with testing data collecting from different time under the same settings (e.g., the distance of the WiFi transceivers, the orientation of the users' chairs and the room of the users). Since nowadays it is common for people who eat alone to sit at a fixed locations, such as at their dining table or office table, we assume that people would not always change the location where they eat every day and we suggest the users rebuild their eating profile with few steps to get a better monitoring if they have to leave their familiar eating places.

\subsection{Impact of Environmental Factors}
Our system is suitable to provide personalized eating monitoring scenario, however, recognizing multiple people at the same time is a common problem in WiFi-based activity monitoring, and other activities from surrounding people can distort the WiFi signal with respect to the eating behavior of one person. However, based on the Fresnel Zone theory, if the distance between the user and surrounding people is longer than the corresponding radius of Fresnel Zone, the impact of people nearby can be negligible. We further conduct a set of experiments to explore the interference of environmental factors from other people with different distances to validate our analysis. Specifically, we asked the users to perform experiments and asked another participant to sit nearby and eat at the same time. Figure XX illustrates the cases of different distances. They are 2 meters and 3 meters. We notice that the performance of the system is impacted little when the distance is 3 meters, which is XX\% lower than the case without any people nearby. However, when the distance is decreased to 2 meters, the interference from other people is more obvious and can not be ignored. According to the results, our system needs the users to collect their training data without much interference, and it is still a challenging task to eliminate the frequency impact of other people moving nearby.

\subsection{A Useful Tools for Potential Research}

%% file: text_infocom/conclusion.tex
\vspace{-2mm}
\section{Conclusion}
In this paper, we explore the feasibility of using the WiFi-enabled devices to provide the users with automatic eating monitoring. We show that the channel state information extracted from the user's smartphone or IoT devices could be utilized to both recognize the user's detailed eating motions with different utensils and detect the minute facial muscle movements of chewing and swallowing, which are essential for inferring when, what and how much the user eats. We develop a device-free system to distinguish eating activities from non-eating based on a K-means cluster methods and then adopt a soft decision-based learning approach to classify the eating motions according to the utensils used by the user. Moreover, we reconstruct the minute facial muscle movements based on the CSI and develop the accumulated power spectral density method to derive the chewing and swallowing statistics. Extensive experiments involving 20 people over 1600-minute eating period are conducted. The results show that the proposed system can achieve up to $95\%$ accuracy for identifying users' eating motions and $10\%$ percentage error for chewing and swallowing amount estimation. 